\documentclass[12pt,a4paper]{article}

\usepackage{relsize}

\usepackage{amssymb}
\usepackage{amsmath,amsfonts}
\usepackage{graphicx}
\usepackage{axodraw}
\usepackage{pstricks}
\usepackage{floatflt}
\usepackage{cite}
\usepackage{wasysym}
\usepackage{hyperref}
\usepackage{graphics}
\usepackage{url}
\usepackage{amssymb}

\newcommand{\nc}{\newcommand}

\nc{\beq}{\begin{equation}}
\nc{\eeq}{\end{equation}}
\nc{\bea}{\begin{eqnarray}}
\nc{\eea}{\end{eqnarray}}
\nc{\nn}{\nonumber}

\nc{\bi}{\begin{itemize}} 
\nc{\ei}{\end{itemize}}

\nc{\veps}{\varepsilon}
\nc{\eps}{\epsilon}
\nc{\as}{\alpha_s}
\nc{\cd}{\cdot}
\nc{\lag}{\cal L }
\nc{\matx}{\left|\cal {M}\right|^2}
\nc{\lqcd}{\Lambda_\textrm{QCD}}
\nc{\msbar}{\overline {\textrm{MS}}}
\nc{\really}{\stackrel{!}{=}}
\def\sla#1{\ifmmode%
\setbox0=\hbox{$#1$}%
\setbox1=\hbox to\wd0{\hss$/$\hss}\else%
\setbox0=\hbox{#1}%
\setbox1=\hbox to\wd0{\hss/\hss}\fi%
#1\hskip-\wd0\box1 }
\nc{\dsla}{\sla{\partial}}
\nc{\Dsla}{\sla{D}}

\def\cf{{\sl c.f.}~}
\def\eg{{\sl e.g.}~}
\def\ie{{\sl i.e.}~}
\def\fig{Fig.~}

\def\eq{Eqn.~}       

\newlength{\nseparation}
\newenvironment{nfigure}[1]
        {\begin{figure}[#1]\hrule\vspace{\nseparation}\par}
        {\vspace{\nseparation}\par \hrule \end{figure}}

\nc{\ib}{\item [$\bullet$]}
\nc{\cds}{\cdots}
\nc{\pr}{\prime}
\nc{\tz}{\tilde{z}}
\def\eg{{\sl e.g.}~}

\nc{\med}{\medskip}

\def\eqs{Eqns.~}

\newbox\charbox
\newbox\slabox
\def\s#1{{      
        \setbox\charbox=\hbox{$#1$}
        \setbox\slabox=\hbox{$/$}
        \dimen\charbox=\ht\slabox
        \advance\dimen\charbox by -\dp\slabox
        \advance\dimen\charbox by -\ht\charbox
        \advance\dimen\charbox by \dp\charbox
        \divide\dimen\charbox by 2
        \raise-\dimen\charbox\hbox to \wd\charbox{\hss/\hss}
        \llap{$#1$}
}}

\newcommand{\eqn}{equation}

\newcommand{\lb}{\left(}
\newcommand{\rb}{\right)}

\newcommand{\al}{\alpha}
\newcommand{\M}{\mathcal{M}}
\newcommand{\mO}{\mathcal{O}}
\newcommand{\ora}{\overrightarrow}

\newcommand{\vareps}{\varepsilon}
\newcommand{\D}{\mathcal{D}}

\newcommand{\ph}{\hat{p}}

\newcommand{\ymax}{y_\text{max}}

\begin{document}
\bibliographystyle{unsrt}
\thispagestyle{empty}
\def\thefootnote{\fnsymbol{footnote}}
\setcounter{footnote}{1}
\null
\vskip 0cm
\begin{center}

{\Large \boldmath{\bf
   Nagy-Soper subtraction scheme\\ for multiparton final states}
    \par} \vskip 2.5em {\large
{\sc Cheng-Han Chung}\\[2ex]
{\normalsize \it Supercomputing Research Center,
National Cheng Kung University,\\ Tainan 701, Taiwan
}\\[2ex]

{\sc Tania Robens}\\[1ex]
{\normalsize \it IKTP, TU Dresden, Zellescher Weg 19, 01069 Dresden, Germany}}
\par \vskip 2em
\end{center}\par

\noindent{\bf Abstract:}\\[0.25em]
\noindent 
In this work, we present the extension of an alternative subtraction scheme for next-to-leading order QCD calculations to the case of an arbitrary number of massless final-state partons. The scheme is based on the splitting kernels of an improved parton shower and comes with a reduced number of final state momentum mappings. While a previous publication including the setup of the scheme has been restricted to cases with maximally two massless partons in the final state, we here provide the final state real emission and integrated subtraction terms for processes with any number of massless partons.  We apply our scheme to three jet production at lepton colliders at next-to-leading order and present results for the differential C parameter distribution.

\par
\null
\setcounter{page}{0}
\clearpage
\def\thefootnote{\arabic{footnote}}
\setcounter{footnote}{0}

\section{Introduction}
With the start of data taking at the LHC in 2009 and its more than successful physics program since then, particle physics has entered an exciting era. Major tasks of the LHC experiments are the accurate measurement of the parameters of the Standard Model (SM) of particle physics, as well as the search for physics beyond the SM. For both, a precise understanding of the SM signals and background processes in an hadronic environment are crucial. These processes are mainly governed by strong interactions, where leading order (LO) calculations can exhibit uncertainties up to $100\,\%$ (\cf \cite{Maestre:2012vp} for a recent review); therefore, for a correct theoretical prediction of these processes at least next-to-leading order (NLO) corrections need to be taken into account. Furthermore, it is generally not sufficient to apply an overall NLO rescaling K-factor to the leading order predictions, as NLO corrections can vary widely for different regions of phase space. More accurate predictions therefore call for the inclusion of these NLO calculations in Monte Carlo event generators, which provide predictions for fully differential corrections to the LO process. Many such generators exist at parton level \cite{mcfm,nlojetpp,Arnold:2008rz,Arnold:2011wj,Ellis:2005zh,Ellis:2006ar,Bern:2011ep,Ita:2011wn,Berger:2010zx,Berger:2009zg,Berger:2009ep,Greiner:2011mp,Binoth:2009rv,Bevilacqua:2011xh,Bevilacqua:2011aa,Bevilacqua:2010qb,Bevilacqua:2009zn}, and in recent years a lot of progress has equally been made to automatize the matching of these processes with parton showers in the Powheg\cite{Frixione:2007vw,Alioli:2008gx,Alioli:2008tz,Papaefstathiou:2009sr,Alioli:2009je,Nason:2009ai,Alioli:2010xd,Hoche:2010pf,Re:2010bp,Alioli:2010qp,Alioli:2010xa,Oleari:2011ey,Melia:2011tj,Alioli:2011nr,Bagnaschi:2011tu,Barze:2012tt,Bernaciak:2012hj,Campbell:2012am,Re:2012zi,Klasen:2012wq,Frederix:2012dh,Jager:2012hd} and (a)MC@NLO \cite{Frixione:2005vw,Papaefstathiou:2009sr,Weydert:2009vr,Torrielli:2010aw,Frixione:2010ra,Frixione:2010wd,Toll:2011tm,Frederix:2011ss,Frederix:2011ig,Hoeche:2012ft} frameworks\footnote{Recent reviews on this can be found in \cite{Hoeche:2011fd,Nason:2012pr}.}.\\

In this paper, we present the generic extension of an improved subtraction scheme\cite{Robens:2010zr,Chung:2010fx,Robens:2011bz}, which facilitates the inclusion of infrared (IR) NLO divergences originating from different phase space contributions in Monte Carlo event generators. These divergences arise whenever internal loop momenta approach zero or particles become collinear, and are known to cancel in any fixed order in perturbation theory \cite{Kinoshita:1962ur, Lee:1964is}. However, these cancellations occur in the combined sum of virtual and real emission contributions, and therefore originate from phase spaces with a different number of particles in the final state. In analytic or semi-numerical calculations, the singularities can be parametrized by an infinitesimal regulator; in the sum of real and virtual contributions, these regulators can then be set to zero to obtain a completely finite prediction. In numerical implementations, however, the inclusion of infinitesimal regulators can easily lead to numerical instabilities. Subtraction methods\cite{Ellis:1980wv,Frixione:1995ms,Catani:1996vz,Dittmaier:1999mb,Denner:2000bj,Weinzierl:2003ra,Phaf:2001gc,Catani:2002hc,Czakon:2009ss, Goetz:2012uz} circumvent this problem by introducing local counterterms that mimic the behaviour of the real emission matrix elements in the singular limits. The integrated counterparts of these terms are then added to the virtual contributions, where again an infinitesimal regulator is used to parametrize the singularities. Then, the higher order contributions in both phase space integrations are finite, and the regulator can be set to zero. In recent years, many of these schemes have been made available on a (semi)automated level  \cite{Gleisberg:2007md, Czakon:2009ss,Hasegawa:2009tx, Frederix:2009yq, Frederix:2010cj, Hoche:2010pf}.\\

While the behaviour of the subtraction terms in the singular limit is determined by factorization \cite{Altarelli:1977zs, Bassetto:1984ik,Dokshitzer:1991wu}, the finite parts of the local counterterms as well as the mapping prescription between real emission and leading order phase space kinematics in the subtraction terms can differ.
Unfortunately, standard schemes \cite{Catani:1996vz,Catani:2002hc}
suffer from a rapidly rising number of momentum mappings, which scales
like $N^{3}$ for a leading order $2\,\rightarrow\,N$
process. Therefore, increasing the number of final state particles
leads to a rapidly rising number of reevaluations of the Born matrix
element. In \cite{Robens:2010zr,Chung:2010fx,Robens:2011bz}, we
therefore proposed a new subtraction scheme with a modified momentum
mapping \cite{Nagy:2007ty,Nagy:2008ns,Nagy:2008eq}, where the number
of momentum mappings scales as $\sim\,N^2$. The momentum mappings are
constructed such that they take the whole remaining event as a
spectator, and the subtraction terms are derived from the splitting
functions in an improved parton
shower\cite{Nagy:2007ty,Nagy:2008ns,Nagy:2008eq}.  In
\cite{Robens:2010zr,Chung:2010fx,Robens:2011bz}, we presented the
scheme for the simplest cases with maximally two partons in the final
state\footnote{Some results for the generic scheme were already
  presented in \cite{Robens:2010zr}.}. In the present work, we extend
the scheme to cases with an arbitrary number of massless particles in
the final state. We also provide the helicity dependent squared
splitting functions for splittings where the mother parton is a
gluon. We validate our scheme by applying it to three-jet production
at NLO at lepton colliders, obtaining complete agreement with the Catani Seymour scheme.\\

This paper is organized as follows. In Section 2, we briefly review the generic setup for subtraction schemes. In Section 3, we review the ingredients of the new scheme and present the generalized results for the integrated subtraction terms for an arbitrary number of final state massless partons. We discuss the application of our scheme to three-jet production at lepton colliders in Section 4. Conclusion and outlook are presented in Section 5. The Appendix contains a summary of the final state splitting functions \cite{Nagy:2007ty} used as subtraction terms, a generic parametrization of four-parton phase space, and the collinear subtraction terms for processes with incoming hadrons.

\section{General structure of NLO cross sections and subtraction schemes}\label{sec:gen_struct}
In this section, we briefly review the general subtraction procedure for calculating NLO cross sections  
at lepton and hadron colliders. We start with a generic cross section at NLO
\begin{eqnarray}
\sigma
&=& \underbrace{\int_m d\sigma^B}_{\sigma^{\text{LO}}}\,+\,\underbrace{\int_{m+1}d\sigma^R\,+\,\int_m d\sigma^V}_{\sigma^{\text{NLO}}}
\end{eqnarray}
where $\sigma$ should be specified by the respective jet
function as discussed below, and $d\sigma^B,\,d\sigma^R,$ and $d\sigma^V$ are the Born, real emission, and virtual contribution respectively. We here consider processes with $m$ particles in the Born-contribution and $m+1$ partons in the real emission terms. After UV-renormalisation, the virtual and
real-emission cross sections each contain infrared and collinear
singularities. These cancel in the sum
of virtual and real contributions\cite{Kinoshita:1962ur, Lee:1964is}, but the individual pieces are
divergent and can therefore not be integrated numerically in four
dimensions.

Subtraction schemes consist of local counterterms that match
the behaviour of the real-emission matrix element in the soft and
collinear regions, and their integrated counterparts. Subtracting these counterterms from the
real-emission matrix elements and adding back the integrated counterparts to the virtual contribution
results in finite integrands for both the virtual correction
($m$-particle phase space) and the real contribution ($(m+1$)-particle
phase space): 
\beq
\label{countertermfinite71335}
d\sigma^{\text{NLO}}\,=\, \left[d\sigma^R-d\sigma^A\right] \,+\,\left[d\sigma^A\,+\,d\sigma^V\right].
\eeq 
The construction of the local counterterms, collectively denoted by
$d\sigma^A$, relies on the
factorisation of the real-emission matrix element in the singular (\ie
soft and collinear) limits (\fig\ref{Dipolefactorizationprocedure_CS})
\cite{Altarelli:1977zs, Bassetto:1984ik,Dokshitzer:1991wu}, and we symbolically write 
\begin{nfigure}{tbp} 
\vspace{-15mm}
\hspace{-25mm}
\includegraphics{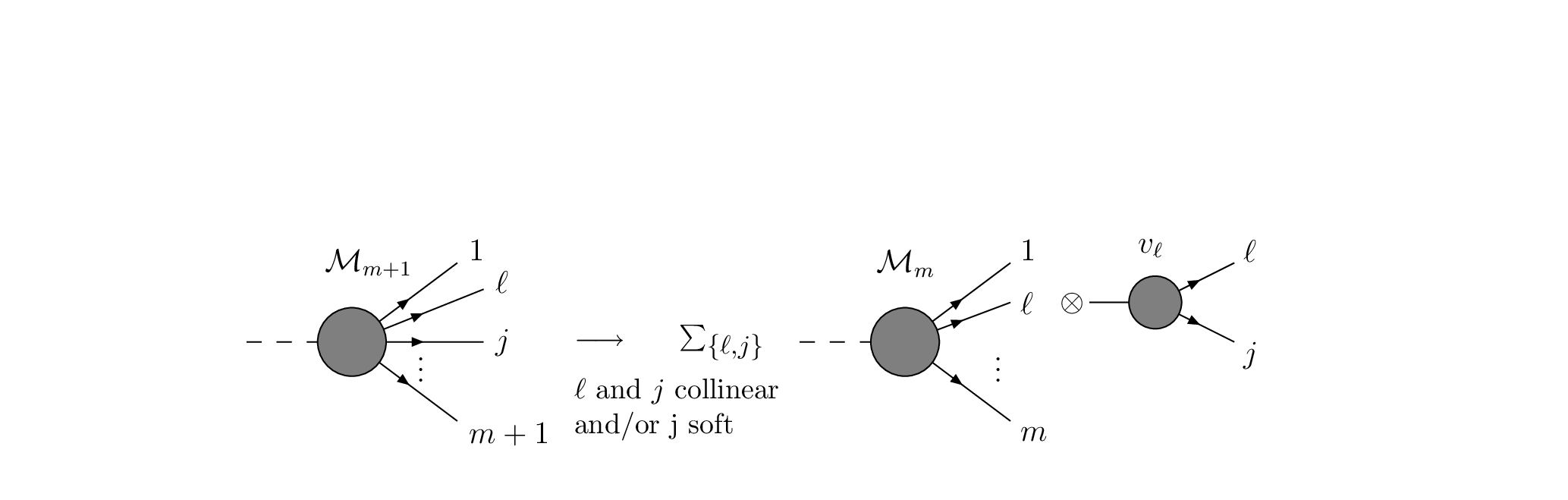}
\caption{Soft/collinear factorisation: when the partons $\ell$ and $j$
  become collinear and/or parton $j$ becomes soft, the ($m+1$)-parton
  matrix element factorises into a sum over $m$-parton matrix elements
  times a singular factor ${v}_{\ell}$.}
\label{Dipolefactorizationprocedure_CS}
\vspace*{0.5cm}
\end{nfigure}

\beq
\label{Factorizationm1toDim}
\left|{\cal M}_{m+1}(\hat p)\right|^2 \quad\longrightarrow\quad \D_\ell\,\otimes\,\left| {\cal M}_{m}( p) \right|^2,
\eeq
where $\D_\ell$ are the dipoles containing the respective singularity structure, and the symbol $\otimes$ denotes a correct convolution in color, spin, and flavour space. $\hat p/\, p$ represent momenta in $(m+1)/\,m$-parton phase space, respectively. As $\left|{\cal M}_{m+1}\right|^2$ and 
$\left| {\cal M}_{m} \right|^2$ live in different
phase spaces, a mapping of their momenta needs to be introduced, which is defined by a mapping function $F_{\textrm{map}}$ according to
\beq\label{eq:fmap}
{p}\,=\,F_{\textrm{map}}\,(\hat p).
\eeq
$\D_\ell$ and its one-parton integrated counterpart $\mathcal{V}_\ell$ are related by 
\beq
\mathcal{V}_\ell\,=\, \int\,d\xi_p \,\D_{\ell},
\eeq
where $d\xi_p$ is an unresolved one parton integration measure.\\

In summary, 
any subtraction scheme needs to fulfill the following requirements:
\begin{itemize}
\item The dipole subtraction terms $\D_\ell$ must match the behaviour of
      the real emission matrix element in each soft and collinear region, and lead to correct IR poles when carrying out the 
      analytical integration over the one parton phase space in a suitable regularization scheme that are necessary to cancel 
      the soft singularities in the virtual (one-loop) matrix element,
\item The mapping function $F_{\textrm{map}}$ guarantees total energy
      momentum conservation as well as the on-shell condition for all external
      particles before and after the mapping.
\end{itemize}
Integrating \eq (\ref{countertermfinite71335}) over phase space using dimensional regularization\cite{Leibbrandt:1975dj,Collins:1984xc}, where $D\,=\,4-2\,\vareps$, then yields
\bea
\label{countertermfinite85}
\sigma^{\text{NLO}}&=&\underset{ \textrm {finite} }
{\underbrace{\int_{m+1}\left[
d\sigma^R-d\sigma^A\right]}}+\underset{ \textrm {finite} }
{\underbrace{\int_{m+1}\,d\sigma^A+\int_m\,d\sigma^V}}                  \nn\\
&=&\int_{m+1}\left[ d\sigma^R_{\vareps=0}-d\sigma^A_{\vareps=0}\right]+
\int_{m}\left[\int_1\,d\sigma^A+d\sigma^V\right]_{\vareps=0}.
\eea
Both integrands are now finite: the integration in $(m+1$) particle
phase space can safely be performed in $D\,=\,4$ dimensions, as the
singular regions are regularized by the respective counterterms. In
the $m$ parton phase space, the sum of the integrated dipole
contribution and the virtual correction does not contain any further
poles, so that we can set $\vareps=0$. Then, all
integrations can be performed numerically. 
The explicit expressions of the cross section $\sigma$ for $m$ and $(m+1$) particle phase space contributions at NLO are
\begin{alignat}{53}
\label{explicitexpressionsNLO}
\int_m\, \left[d\sigma^B\,+\,d\sigma^V\,+\,\int_1\,d\sigma^A\right]& =\int    dPS_m
\left[\left| {\cal M}_{m} \right|^2\,+\,\left| {\cal M}_{m} \right|^2_{\textrm{one-loop}}\,+\,
\sum_\ell\,\mathcal{V}_\ell\,\otimes\,\left| {\cal M}_{m} \right|^2\right], \notag \\
\int_{m+1}\,\left[ d\sigma^R - d\sigma^A \right]&=\int dPS_{m+1} \left[\left| {\cal M}_{m+1} \right|^2 \,-\, \sum_\ell\,D_\ell\,\otimes\,\left| {\cal M}_{m} \right|^2\right],  
\end{alignat}
where in this symbolic notation $\int dPS$ includes all flux and
symmetry factors, and with  $\left| {\cal M}_{m}\right|^2$, $\left| {\cal M}_{m+1}\right|^2$ and 
$\left| {\cal M}_{m}\right|^2_{\textrm{one-loop}}$ being the squared LO
matrix element, the squared real emission matrix element and the
interference term, respectively. In \eq(\ref{explicitexpressionsNLO}), the sum runs
over all local counterterms needed to match the complete singularity
structure of the real emission contribution, and
convolution with jet functions then ensures the collinear and
infrared safety of the Born-level
contribution. The insertion operator $I(\vareps)$ is then defined on a
cross section level according to
\begin{\eqn*}
\int_{m}   \int_{1} d\sigma^A = \int_m  d\sigma^B \otimes { I}(\vareps),
\end{\eqn*}
where the symbol $\otimes$ again denotes a proper convolution in spin,
color, and phase space.
The generalization of this for processes with initial
state hadrons has already been presented in \cite{Chung:2010fx}; for completeness,
we repeat the argument in Appendix \ref{app:cts}.\\

\subsubsection*{Observable-dependent formulation of the subtraction method}
\label{Observableindependentformulation}
The jet observables should be well defined such that
the leading order cross sections are infrared and collinear safe. 
The jet cross sections are defined as
\begin{alignat}{53}
\sigma^{ LO}_J&= \int dPS_m (p_1,\cdots,p_m)\, \left| {\cal M}_{m} (p_1,\cdots,p_m)\right|^2\, F_J^{(m)} (p_1,\cdots,p_m),                                        \notag \\
\sigma^{ NLO}_J&=\int dPS_{m+1} (p_1,\cdots,p_{m+1})\,\left| {\cal M}_{m+1} (p_1,\cdots,p_{m+1})\right|^2 \,
F_J^{(m+1)} (p_1,\cdots,p_{m+1})     \notag \\
&+ \int dPS_{m} (p_1,\cdots,p_m)\,\left| {\cal M}_{m} (p_1,\cdots,p_m)\right|^2_{\textrm{one-loop}}\, 
F_J^{(m)} (p_1,\cdots,p_m). 
\end{alignat}
In general, the jet
function  may contain $\theta$ functions (which define cuts and
corresponding  cross sections) and $\delta$ functions (which define
differential cross sections). For an infrared finite jet function, we require that
\begin{eqnarray}\label{jetobservableinfraredandcollinearsafe}
F_J^{(m+1)} (p_1, \cdots ,p_j=\lambda\,q, \cdots ,p_{m+1} ) &\to& F_J^{(m)} (p_1,\cdots ,p_{m+1} ), 
                       \quad {\textrm {if}}\quad \lambda \to 0                                    \nn    \\
F_J^{(m+1)} (p_1,..,p_i,..,p_j,..,p_{m+1} ) &\to& F_J^{(m)} (p_1,..,p,..,p_{m+1} ) 
                                      \nn   \\ 
& & \quad \quad \quad \quad {\textrm {if}}\quad   p_i\to z\,p,\,  p_j\to (1-z)\,p,  \nn   \\
F_J^{(m)} (p_1,\cdots ,p_m ) &\to& 0,   \quad{\textrm {if}}\quad  p_i \cdot p_j \to 0.  
\label{jetobservableinfraredandcollinearsafe}
\end{eqnarray}
The last condition of
\eq(\ref{jetobservableinfraredandcollinearsafe}) corresponds to an infrared safe definition of the Born-level observable, while the first two conditions guarantee infrared and collinear safety of the observables and can be summarized to
\beq\label{eq:fjcond}
F_J^{(m+1)}\,\to \,F_J^{(m)}
\eeq
in the singular limits.\\

\noindent
We then have
\begin{eqnarray}\label{eq:jetincl}
\lefteqn{\int_m\, \left[d\sigma^B\,+\,d\sigma^V\,+\,\int_1\,d\sigma^A\right]}\nn \\
& =&\int    dPS_m
\left[\left| {\cal M}_{m} \right|^2\,+\,\left| {\cal M}_{m} \right|^2_{\textrm{one-loop}}\,+\,
\sum_\ell\,\mathcal{V}_\ell\,\otimes\,\left| {\cal M}_{m} \right|^2\right]\,  F_J^{(m)}(p), \nn \\
\lefteqn{ \int_{m+1}\,\left[ d\sigma^R - d\sigma^A \right]}\nn \\&=&\int dPS_{m+1}\, \left[\left| {\cal M}_{m+1} \right|^2\, F_J^{(m+1)}(\ph) \,-\, \sum_\ell\,\D_\ell\,\otimes\,\left| {\cal M}_{m} \right|^2 F_J^{(m)}(p) \right],  \nn\\
&& 
\end{eqnarray}
where in the integrated subtraction term the momenta $p$ are derived from $\ph$ using the respective momentum mapping.

\section{Alternative subtraction scheme: setup}
In this section, we will first review the setup of our scheme as well as the mapping and respective
integration measures which have already been presented in \cite{Nagy:2007ty,Chung:2010fx}.
In our scheme, the NLO subtraction terms are
derived from the splitting functions introduced in a parton shower context \cite{Nagy:2007ty,Nagy:2008ns,Nagy:2008eq}, and the
$m+1$ to $m$
phase space mappings needed correspond to the inverse of the
respective shower $m$ to $m+1$ mappings. In the following, we will
denote the $m+1$ phase space four-vectors by $\ph_{1},\ph_{2},.. .$ and
$m$ phase space four-vectors by $p_{1},\,p_{2},\,...$.  In $m+1$ phase space, the four-momenta of the emitter, emitted particle, and spectator are denoted $\ph_{\ell},\,\ph_{j},$ and $\ph_{k}$ respectively. Note that here the spectator needs to be
specified only if $\ph_{j}$ denotes a gluon, as
  we use the whole remaining event as a spectator in the sense of
  momentum redistribution for both initial and final state
  mappings. We here restrict our expressions to
subtractions on the parton level and to massless partons.

\subsection{Splitting functions}\label{sec:splitf}
We start with a description of the matrix element factorization in the soft and collinear limits, following the notation in \cite{Nagy:2007ty}, where the QCD scattering amplitude for $m+1$ partons is given as a vector in (colour $\otimes$ spin) space, 
\beq
\mid {\cal M}(\{\hat p, \hat f\}_{m+1})\rangle.
\eeq
In the singular limits, the amplitude $\mid {\cal M}_\ell(\{\hat p, \hat f\}_{m+1})\rangle$ can be factorized into a splitting 
operator times the $m$-parton matrix element 
\beq
\label{QCDFactorizationm1tVm}
\mid {\cal M}_\ell(\{\hat p, \hat f\}_{m+1})\rangle\,=\,t^\dagger_\ell(f_\ell \to \hat f_\ell + 
\hat f_{j})\,V^\dagger_\ell(\{\hat p, \hat f\}_{m+1})\,\mid {\cal M}(\{ p,  f\}_{m})\rangle,
\eeq
where the index $\ell$ labels the emitter/ mother parton in the $(m+1$)/
$m$ particle phase space. $V^\dagger_\ell(\{\hat p, \hat f\}_{m+1})$ is an operator acting on the spin part of the (colour $\otimes$ spin) space, while 
$t^\dagger_\ell(f_\ell \to \hat f_\ell + \hat f_{j})$ is an operator acting on the colour part of the (colour $\otimes$ spin) space.
The Born amplitude for producing $m$ partons is evaluated at momenta and flavours $\{p, f\}_{m}$ determined from 
$\{\hat p, \hat f\}_{m+1}$ according to the respective momentum mappings.
The spin-dependent splitting operator can be described in the spin space $\mid \{s\}_m\rangle$:
\beq
\label{hatsVdaggers}
\langle \{\hat s\}_{m+1}\mid V^\dagger_\ell(\{\hat p, \hat f\}_{m+1})\mid \{s\}_m\rangle.
\eeq
If we take \eq (\ref{hatsVdaggers}) to be diagonal, we can define the splitting functions $v_{\ell}$ according to
\begin{alignat}{53}
\langle \{\hat s\}_{m+1}\mid V^\dagger_\ell(\{\hat p, \hat f\}_{m+1})\mid \{s\}_m\rangle\,=\,
\left(\prod_{n\notin\{\ell,j=m+1\}} \delta_{\hat s_n,s_n}\right)\,
v_\ell (\{\hat p, \hat f\}_{m+1},\hat s_{j},\hat s_{\ell},s_\ell).
\label{eq:vel_def}
\end{alignat}
Explicit forms for the splitting functions $v_{\ell}$ have been presented in \cite{Nagy:2007ty}. For the construction of the subtraction terms, we consider the
approximation for the squared matrix element in the singular limits 
\begin{\eqn*}
\sum_{\ell,\ell'}\,\langle {\cal M}_\ell(\{\hat p, \hat f\}_{m+1}|{\cal M}_{\ell'}(\{\hat p, \hat f\}_{m+1}\rangle\,\sim\,\sum_{\ell,\ell'}\,\,v^{*}_{\ell}v_{\ell'} \langle{\cal M}(\{ p_{\ell},  f\}_{m}) |{\cal M}(\{ p_{\ell'},  f\}_{m})  \rangle.
\end{\eqn*} 
For the direct splitting function, where $\ell\,=\,\ell'$, we obtain
\beq\label{eq:Wellell}
W_{\ell\ell}\,\equiv\,v^{2}_{\ell}\,=\,v_\ell (\{\hat p, \hat f\}_{m+1},\hat s_{j},\hat s_{\ell},s_\ell)\,
v^*_\ell (\{\hat p, \hat f\}_{m+1},\hat s_{j},\hat s_{\ell},s_\ell),
\eeq
which, after summing over the daughter parton spins and averaging over the mother parton spins, leads to the spin-averaged splitting functions $\overline{W}_{\ell \ell}$ as subtraction terms. If the mother parton is a gluon, the Born-type matrix element might have an explicit dependence on the gluons polarization; in this case, we need to use
\begin{\eqn}\label{eq:spincorr}
\langle \nu|W_{\ell \ell}|\nu'\rangle
\end{\eqn}
in the real-emission subtraction terms, where $\nu,\,\nu'$ are the polarisation indices of the $m$-parton phase space gluon. If the spin correlation tensor defined by Eqn. (\ref{eq:spincorr}) is perpendicular to $p_\ell$, the angular correlations vanish after the integration over the unresolved particles phase space and the integral over $\overline{W}_{\ell \ell}$ still provides the correct integrated counterterm \cite{Catani:1996vz}.
For the collinear terms, the colour factors can easily be obtained \cite{Nagy:2007ty}:
\begin{eqnarray*}
C(\hat{f}_{\ell},\hat{f}_{j})\,=\,
\begin{cases}
C_{F}& (\hat{f}_{\ell},\hat{f}_{j})\,=\,(q,g), (g,q),\\
C_{A}& (\hat{f}_{\ell},\hat{f}_{j})\,=\,(g,g),\\
T_{R}& (\hat{f}_{\ell},\hat{f}_{j})\,=\,(q,\bar{q}).\\
\end{cases}
\end{eqnarray*}
For soft gluon emissions, we also have to consider terms for which $\ell\,\neq\,\ell'$, which we will describe below.

\subsubsection{Eikonal factor} 
When a gluon with four-vector $\ph_{j}$ becomes soft, or soft and collinear with $\ph_\ell$, the splitting amplitude $v_\ell$ defined in \eq (\ref{eq:vel_def}) can be replaced by the eikonal approximation for $\ph_{j}\,\rightarrow\,0$
\beq\label{eq:veik}
v_\ell^{\textrm {eik}} (\{\hat p, \hat f\}_{m+1},\hat s_{j},\hat s_{\ell},s_\ell) \,=\,
   \sqrt{4\pi\as}\,\delta_{\hat s_\ell, s_\ell} \,
   \frac{\veps (\hat p_j,\hat s_j,\hat Q)^*\cd \hat p_\ell}
   {\hat p_j\cd \hat p_\ell},
\eeq
where $\veps (\hat p_j,\hat s_j,\hat Q)$ denotes the polarization vector of the emitted gluon with spin $s_j$. $\hat{Q}$ denotes the total momentum of the $(m+1)$ phase space event and is used as a gauge vector. 
The eikonal approximation of the spin-averaged splitting functions $\overline W_{\ell\ell}$ is then
\beq
\label{spinaveragedsplittingfunctions_Wll_eikonal}
\overline{W}_{\ell\ell}^{\textrm {eik}}\,=\, 4\,\pi\,\as\, 
\frac{\hat p_\ell\cd  D(\hat p_{j},\hat Q) \cd  \hat p_\ell}
     { (\hat p_{j}\cd \hat p_\ell)^2 },
\eeq
where flavour-dependent averaging factors are already taken into account. 
The transverse projection tensor $D^{\mu\nu}$ is given by
\beq
\label{transverseprojectiontensor}
D^{\mu\nu}(\hat p_j,\hat Q) \,=\,- g^{\mu\nu} + \frac{\hat p_j^\mu\, \hat Q^\nu + \hat Q^\mu\, \hat p_j^\nu}
{\hat p_j\cd \hat Q}- \frac{\hat Q^2\, \hat p_j^\mu\, \hat p_j^\nu}{(\hat p_j\cd \hat Q)^2}.\nn
\eeq
It will be convenient to define a dimensionless function $F$:
\beq
F\,=\,\frac{\hat p_\ell\cdot\hat p_j}{4\,\pi\,\as}\, \overline W_{\ell\ell}.\nn
\eeq
We then have
\beq
F_{\textrm{eik}} \,\equiv \,  \frac{\hat p_\ell\cdot\hat p_j}{4\,\pi\,\as}\, \overline W_{\ell\ell}^{\textrm{eik}} \,=\,
\frac{\hat p_\ell\cdot  D(\hat p_j, \hat Q)\cdot\hat p_\ell}{\hat p_\ell\cdot\hat p_j}\,=\, \frac{2\,\hat p_\ell\cdot Q}
{\hat p_j\cdot Q}-\frac{ Q^2\,\hat p_\ell\cdot\hat p_j}
{(\hat p_j\cdot Q)^2}. \nn
\eeq
The eikonal factor, in combination with the interference terms, is then used to construct dipole partitioning functions. 
\subsubsection{Soft splitting functions}\label{sec:softss}
\label{Softsplittingfunction2242010}
\begin{nfigure}{tbp} 
\vspace{-15mm}
\hspace{-25mm}
\includegraphics{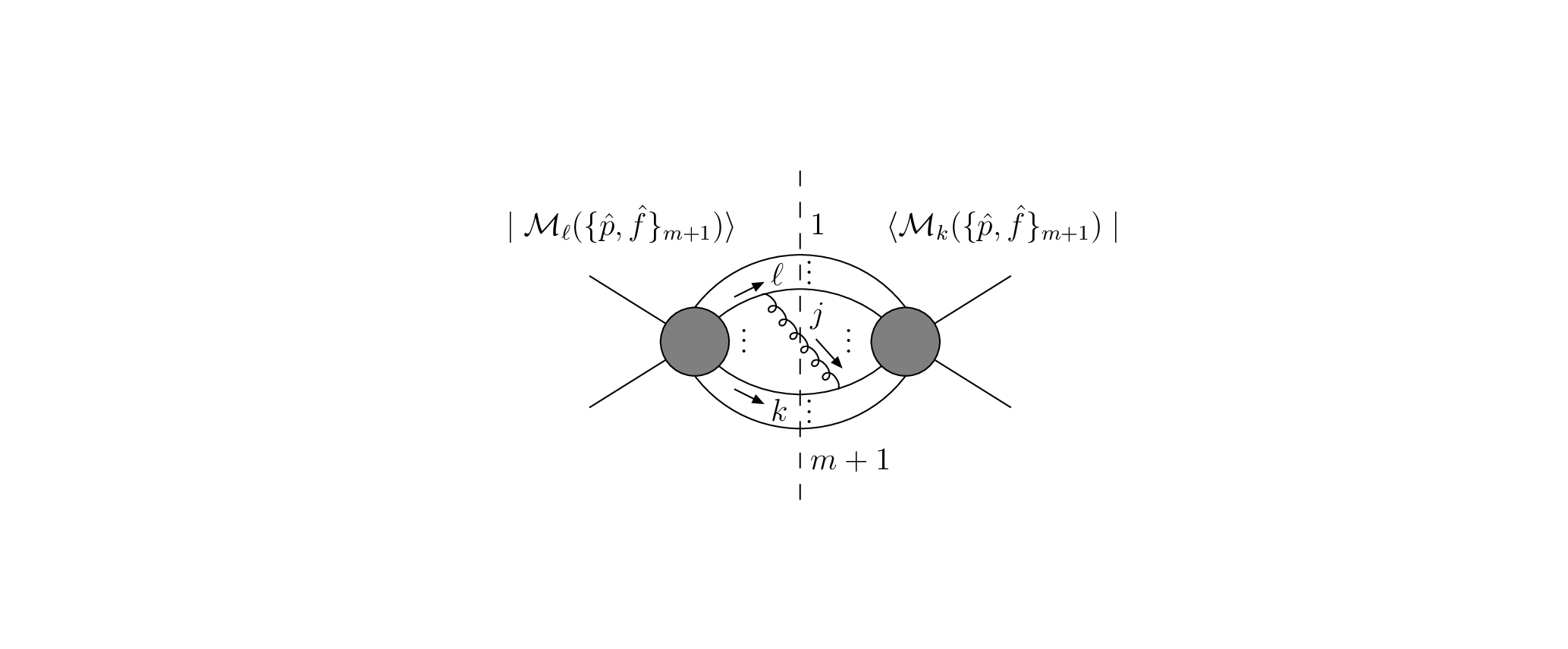}
\vspace{-25mm}
\caption{Soft diagram:  parton $j$ is emitted from parton $\ell$ in the scattering amplitude and parton $j$ is emitted from
   parton $k$ in the complex-conjugate scattering amplitude. }     
 \label{amplitudesquare_Softdiagram_qqg}
\vspace*{0.7cm}
\end{nfigure}
For soft gluon emission, we also need to take interference diagrams between different
emitters into account. This means the emitted parton $j$ can be emitted from emitter $\ell$ in the amplitude and parton $j$ can also be emitted from a
different emitter $k$ in the complex-conjugate amplitude (\fig\ref{amplitudesquare_Softdiagram_qqg}). The interference splitting function is then given by
\beq
\label{interferencesplittingfunctionellk317285}
\overline{W}_{\ell k}\,\sim\,
v_\ell(\{\hat p, \hat f\}_{m+1}, \hat s_j, \hat s_\ell, s_\ell)\,
   v_k(\{\hat p, \hat f\}_{m+1}, \hat s_j, \hat s_k,    s_k)^* \,
    \delta_{\hat s_\ell, s_\ell} \,
    \delta_{\hat s_k, s_k}.
\eeq
The splitting function \eq(\ref{interferencesplittingfunctionellk317285}) contains a singularity when the emitted gluon $j$ is soft; however 
when gluon $j$ is collinear with parton $\ell$ or $k$, it does not contribute to a leading singularity. 
In the special case that $\hat p_{j}$ is soft, or possibly soft and collinear with $\hat p_\ell$, we can use:
\beq
\overline{W}_{\ell k}\,\sim\,
v_\ell^{\textrm {eik}}(\{\hat p, \hat f\}_{m+1}, \hat s_j, \hat s_\ell, s_\ell)\,
   v_k^{\textrm {eik}}(\{\hat p, \hat f\}_{m+1}, \hat s_j, \hat s_k,    s_k)^* \,
    \delta_{\hat s_\ell, s_\ell} \,
    \delta_{\hat s_k, s_k}.\nn
\eeq
Note that this term contributes only if particle $j$ is a gluon.
In this prescription, there is an ambiguity in the allocation of the
singularities, which can be distributed with the help of dipole
partitioning functions; for completeness, we here repeat the argument
in \cite{Nagy:2007ty,Nagy:2008ns}. The complete sum over all singular terms will contain a term
\beq
W_{\ell k,k\ell}\,=\,W_{\ell k}\,t^{\dagger}_{\ell}\,\otimes t_{k}\,+\,W_{k \ell}\,t^{\dagger}_{k}\,\otimes t_{\ell}. \nn
\eeq   
For each of the two contributions, we can now introduce weight factors which redistribute the splitting functions to the corresponding mappings
\beq
W_{\ell k}\,\longrightarrow\,A_{\ell k}\,W^{(\ell)}_{\ell k}+A_{k \ell}W_{\ell k}^{(k)},
\eeq
where
\beq
A_{\ell k}\,+\,A_{k \ell}\,=\,1\nn
\eeq
for any fixed momenta. $W^{(\ell)}$ denotes that for the mapping of
this part of the interference term, $\ph_{\ell}$ is considered to be
the emitter; for $W^{(k)}$, particle $\ph_{k}$ acts as the emitter, such that the roles of $\ell$ and $k$ are interchanged.
We then have for the total sum of the two distributions
\beq
W_{\ell k,k \ell}\,=\,A_{\ell k}\,\left[ W^{(\ell)}_{\ell k}\,t^{\dagger}_{\ell}\otimes\,t_{k} \,+\, W^{(\ell)}_{k \ell}\,t^{\dagger}_{k}\otimes\,t_{\ell}\right]\,+\,A_{k\ell}\,\left[ W^{(k)}_{\ell k}\,t^{\dagger}_{\ell}\otimes\,t_{k} \,+\, W^{(k)}_{k\ell}\,t^{\dagger}_{k}\otimes\,t_{\ell}\right].\nn
\eeq
We now combine this with the pure squared splitting function $\overline{W}_{\ell \ell}$ with the colour factor $t^{\dagger}_{\ell}\,\otimes\,t_{\ell}$. Invariance of the matrix element under colour rotations implies \cite{Nagy:2008ns}:
\beq
t^{\dagger}_{\ell}\,\otimes\,t_{\ell}\,=\,-\sum_{k\,\neq\,\ell}\,\frac{1}{2}\,\left[t^{\dagger}_{k}\,\otimes\,t_{\ell}+ t^{\dagger}_{\ell}\,\otimes\,t_{k}\right],\nn
\eeq
and the complete contribution obeying one mapping is then given by
\beq\label{eq:addsubif}
-\frac{1}{2}\,\left[t^{\dagger}_{k}\,\otimes\,t_{\ell}+ t^{\dagger}_{\ell}\,\otimes\,t_{k}\right]\,\left[\overline{W}_{\ell \ell}-\overline{W}_{\ell k} \right]
\eeq
with the spin-averaged interference contribution
\begin{equation}
\label{Wellk2842010}
\overline W_{\ell k} \,=\, 4\,\pi\,\as\, 2\, A_{\ell k}
\frac{\hat p_\ell\cd D(\hat p_{j}, \hat Q) \cd  \hat p_k}
{\hat p_{j}\cd \hat p_\ell \, \hat p_{j}\cd \hat p_k}.
\end{equation}
We now split the collinear and soft parts of the respective spin-averaged splitting functions
according to 
\beq
\label{splitsplittingfunctions335}
\overline{W}_{\ell\ell}- \overline{W}_{\ell k} \,=\,\left( \overline{W}_{\ell\ell} - \overline{W}_{\ell\ell}^{\textrm {eik}} \right)
+ \left(\overline{W}_{\ell\ell}^{\textrm {eik}}  - \overline{W}_{\ell k}\right).
\eeq
The second part of \eq(\ref{splitsplittingfunctions335}) can be expressed in
terms of dipole partitioning functions $A_{\ell k}'$ \cite{Nagy:2008eq}:
\beq
\overline{W}_{\ell\ell}^{\textrm {eik}}  - \overline{W}_{\ell k}\,=\,
4\,\pi\,\as\,A_{\ell k}' \,\frac{-\hat{P}_{\ell k}^{2}} {(\hat p_j  \cd \hat p_\ell \,\hat p_j  \cd \hat p_k )^{2}},\nn
\eeq
where  $ \hat{P}_{\ell k} \,=\,(\hat p_j  \cd \hat p_\ell) \,\hat p_k - (\hat p_j  \cd \hat p_k) \,\hat p_\ell  $. 
Several choices for $A_{\ell k}'$ have been proposed in  \cite{Nagy:2008eq}; all results given here have been obtained using Eqn. (7.12) therein:
\beq
\label{Nagy2008eq712}
A_{\ell k}' (\{\hat p\}_{m+1})\,=\,\frac{\hat p_j  \cd \hat p_k\; \hat p_\ell\cd \hat Q}
{\hat p_j  \cd \hat p_k\; \hat p_\ell\cd \hat Q +\hat p_j  \cd \hat p_\ell\; \hat p_k\cd \hat Q}.\nn
\eeq
The partitioning weight function $A_{\ell k}' $ also obeys the relation $A_{\ell k}'(\{\hat p\}_{m+1}) + A_{k\ell}'(\{\hat p\}_{m+1}) = 1$. 
The general form of the interference spin-averaged splitting function is then given by
\beq
\label{interferencespinaveragedsplittingfunction}
\Delta W_{\ell k} \,=\, \overline{W}_{\ell\ell}^{\textrm {eik}}  - \overline{W}_{\ell k} \,=\,4\,\pi\,\al_s
\frac{2\, \hat p_\ell\cdot\hat p_k\, \hat p_\ell\cdot\hat Q } 
{\hat p_\ell\cdot\hat p_j\,
\left(\hat p_j\cdot\hat p_k\, \hat p_\ell\cdot\hat Q+\hat p_\ell\cdot\hat p_j\,\hat p_k\cdot\hat Q \right)}
\eeq
The corresponding color factor is defined by Eqn. (\ref{eq:addsubif}) as
\begin{\eqn}\label{eq:clk}
C_{\ell\,k}\,\equiv\,-\frac{1}{2}\,\left[t^{\dagger}_{k}\,\otimes\,t_{\ell}+ t^{\dagger}_{\ell}\,\otimes\,t_{k}\right].
\end{\eqn}
The only singularity in \eq(\ref{interferencespinaveragedsplittingfunction}) arises from the factor $\hat p_\ell\cd \hat p_j $ in the denominator; the interference term is constructed such that it vanishes for the collinear singularity from
$ \hat p_j\cd \hat p_k  \,\rightarrow\,0$. We also assume that the variables considered are such that they are finite for
$p_{\ell}\cd p_{k}\,\rightarrow\,0$, \ie singularities arising in this limit should be taken care of by the definition of the jet function as described in Section \ref{sec:gen_struct}. The interference term only needs to be considered if the emitted
parton $j$ is a gluon. If parton $j$ is a quark or antiquark, this
term vanishes.

\subsection{Final state momentum mapping} \label{sec:mommap}
In this section, we will describe the momentum mapping that is used in the shower prescription \cite{Nagy:2007ty,Nagy:2008ns,Nagy:2008eq} as well as the subtraction scheme.
As before, hatted momenta $\left\{\ph_n\right\}$ are used to describe $(m+1)$-parton phase space and unhatted momenta $\left\{p_n\right\}$ $m$-parton phase space particles; emitter, emitted parton and spectator are labeled $\ph_\ell,\,\ph_j,$ and $\ph_k$ respectively. The four-vectors $\ph_a, \ph_b$ refer to initial state partons.\\

For a parton splitting
\begin{\eqn*}
p_\ell\,\rightarrow\,\ph_\ell+\ph_j
\end{\eqn*}
on-shellness of all momenta in both $m$ and $(m+1$) phase space requires a momentum mapping which reduces to
\begin{\eqn*}
p_\ell\,=\,\ph_\ell+\ph_j
\end{\eqn*}
in the singular limits; away from these kinematic regions, an additional spectator momentum needs to be modified to guarantee $p^2_i\,=\,\ph^2_i\,=\,0$ for all particles. In our scheme, we use the whole remaining event as a spectator, which leads to a scaling behaviour $\sim\,N^2/2$ for the number of required mappings,  where $N$ is the number of final state partons in the process\footnote{In the Catani Seymour scheme, each additional parton in the process subsequently serves as a spectator, leading to an overall scaling behaviour $\sim\,N^3/2$ for number of required mappings.}.

\subsubsection{Mapping in the parton shower}
For a final state splitting, we leave the momenta of the initial state partons unchanged:
\beq
p_a\,=\,\hat p_a,\quad p_b\,=\,\hat p_b. \nn
\eeq
Let $Q$ be the total momentum of the final state partons
\beq
Q\, \equiv\, \sum_{n = 1}^m\, p_n \,=\, p_a + p_b.\\
\eeq
Here the momenta of the incoming partons remain the same, hence $Q\,=\,\hat Q\,=\,\hat p_a +\hat p_b$. We define 
\beq
\label{a_ell_parameter}
 a_\ell\,=\,\frac{ Q^2}{2\,p_\ell\cdot Q}, 
\eeq
where $a_\ell \ge 1$. The momenta of the daughter partons $\hat p_\ell$ and $\hat p_j$ are then mapped according to
\beq
\label{totalmomentumfinalstatepellpj}
 P_\ell\,=\,\hat p_\ell+\hat p_j\,=\, \lambda\,p_\ell + \frac{1-\lambda+y}{2\, a_\ell}\,  Q. 
\eeq
The parameters $\lambda$ and $y$ follow from energy
momentum conservation as
\beq\label{eq:ydef}
\lambda\,=\,\sqrt{\left(1+y\right)^2-4\,a_\ell\,y}, \qquad y\,=\,\frac{\hat p_\ell\cdot\hat p_j}{ p_\ell\cdot Q}. 
\eeq
$y$ is a measure for the virtuality of the splitting, with
\beq	
\label{ymax}
y_\textrm{max}\,=\,\left( \sqrt{a_\ell}-\sqrt{a_\ell-1}\,\right)^{2}\,=\,2\,a_\ell\,-\,1\,-2\,\sqrt{a_\ell\,(a_\ell-1)} 
\eeq
and $\lambda(y_\text{max})=0.$\\

The mapping prescription used in our scheme now defines the whole remaining event as a spectator, \ie the momenta of all corresponding final state particles are mapped as
\beq
\label{hatpLambdap}
\hat p^{\mu}_{n}\,=\,\Lambda (\hat{K},  K)^\mu{}_\nu \, p_n^\nu  ,\quad n\notin\{\ell,j=m+1\}
\eeq
with the Lorentz transformation 
\begin{eqnarray}\label{eq:LTini}
&&\Lambda^{\mu\nu} (K_1,K_2)\,=\,g^{\mu\nu}\,-\,\frac{2\,(K_1+K_2)^{\mu}\,(K_1+K_2)^{\nu}}{(K_1+K_2)^{2}}\,+\,\frac{2\,K_1^{\mu}\,K_2^{\nu}}{K_2^{2}}.
\end{eqnarray}
$K$ and $\hat{K}$ are given by 
\bea
K\,=\, Q - p_\ell,\;\hat{K}\,=\, Q - P_\ell,
\eea
and correspond to the total momentum of the final state spectators before and after the splitting respectively, with
\beq 
\hat{K}^\mu \,=\, \Lambda (\hat{K},K)^\mu{}_\nu  \, K^\nu. \nn
\eeq
For $a_\ell\,=\,1$, this simplifies to
\begin{\eqn*}
\hat{K}\,=\,(1-y)\,K
\end{\eqn*}
and we therefore have
\begin{\eqn}\label{eq:lambdaa1}
\Lambda(\hat{K},K)^{\mu\nu} (a_\ell\,=\,1)\,=\,(1-y)\,g^{\mu\nu}.
\end{\eqn}

The flavours of the spectator partons remain unchanged
\beq
\hat f_n\,=\,f_n,\quad n\notin \{\ell, j=m+1\},\nn
\eeq
while the flavour of the mother parton $f_\ell$ obeys   
\beq
\hat f_\ell+\hat f_j \,=\,f_\ell;\nn
\eeq
\eg if the mother parton $\ell$ is a quark/antiquark, then  we have $(\hat f_\ell, \hat f_j)\,=\,(q/\bar q, g)$. If the mother parton $\ell$ is a gluon, then $(\hat f_\ell, \hat f_j)$ can be a pair of gluons $(g, g)$, which corresponds to $g\to g\,g$ splitting, or any choice of quark/antiquark flavours $(q, \bar q)$, which corresponds to $g\to q\, \bar q$ splitting.

\subsubsection{Mapping in the subtraction scheme}\label{sec:mommapsub}

There is an inverse of the above mapping prescription, which maps the $(m+1)$-parton momenta to the $m$-parton momenta needed for the evaluation of the real-emission subtraction terms. We
start with $\{\hat p\}_{m+1}$ and determine $\{p\}_{m}$.
The momentum $p_\ell$ of the mother parton follows directly from \eq(\ref{totalmomentumfinalstatepellpj})
\beq\label{eq:fin_map}
p_\ell\,=\,\frac{1}{\lambda}\,(\hat p_\ell + \hat p_j)-\frac{1 - \lambda + y}{2\, \lambda\, a_\ell}\, Q. 
\eeq
The parameters $y$ and $a_\ell$ read
\beq\label{eq:yinverse}
y = \frac{P_\ell^2}{2\, P_\ell\cdot Q - P_\ell^2}  \quad \mbox{and} \quad 
a_\ell \,=\, \frac{ Q^2}{2\, P_\ell\cdot Q - P_\ell^2}\,, 
\eeq
with $P_\ell \,= \, \hat p_\ell+\hat p_j$. The parameter $\lambda$ then follows from
Eqn.~(\ref{eq:ydef}).\\

Now we need the inverse Lorentz transformation to \eq(\ref{hatpLambdap}), which is used to map all nonemitting final state spectators. We have
\beq
\label{pLambdahatp}
 p_n^\mu\,=\,\Lambda (K,\hat{K})^\mu{}_\nu \,\hat p^{\nu}_{n} ,\quad n\notin\{\ell,j=m+1\},
\eeq
where $\Lambda (K,\hat{K})^\mu{}_\nu$ is given by \eq(\ref{eq:LTini}). For $a_{\ell}=1$, the mapping reduces to
\begin{\eqn}\label{eq:fin_map_a1}
p_\ell\,=\,\frac{1}{1-y}\,\lb P_\ell-y\,Q \rb,\,p_k\,=\,\frac{\ph_{k}}{1-y}.
\end{\eqn}

The flavour transformation is similar to the case of parton splitting. The 
flavour of the mother parton $f_\ell$ is given by 
\beq
f_\ell\,=\, \hat f_\ell+\hat f_j, \nn
\eeq 
with the rule of adding flavours, $q+g=q$ and $q+\bar q =g$. The flavours of the spectators remain unchanged: 
\beq
f_n\,=\,\hat f_n, \quad n\notin \{\ell, j=m+1\}. \nn
\eeq

\subsubsection{Phase space factorization}
In the integration of the subtraction terms over the one-parton
unresolved phase space, we use the generic phase-space factorisation
\begin{\eqn}
\left[d\{\ph,\hat{f}\}_{m+1}\right]g(\{\ph,\hat{f}\}_{m+1})\,=\,
\left[d\{p,f\}_{m}\right]\,d\xi_{p}g(\{\ph,\hat{f}\}_{m+1})\,,
\end{\eqn}
where $g(\{\hat p, \hat f\}_{m+1})$ is an arbitrary function. In this work, we chose to regularise the infrared and collinear singularities that appear in the splitting functions using dimensional regularisation,
\ie we work in $D\,=\,4 - 2\vareps$ dimensions so that the
singularities appear as $1/\vareps^2$ (soft and collinear) and
$1/\vareps$ (soft or collinear) poles. We then have for
the unresolved one-parton integration measure
\begin{alignat}{5}
d\xi_p&=
dy\,\theta(y_\textrm{min}< y  <
y_\textrm{max})\,\lambda^{D-3}\,\frac{p_{\ell}\cdot Q}{\pi}\,
\frac{d^D\hat p_{\ell}}{(2\,\pi)^{D}}\,
2\,\pi\,\delta^{+}(\hat p_{\ell}^{2} )\,
\frac{d^{D}\hat p_{j}}{(2\,\pi)^{D}}\,2\,\pi\,\delta^{+}(\hat p_{j}^{2} )\, 
\notag \\
&\times
(2\,\pi)^{D}\,\delta^{(D)}\left(\hat p_\ell + \hat p_{j}- \lambda\, p_\ell - \frac{1 - \lambda  + y}{2\,a_\ell}\, Q
\right).
\label{eq:fin_meas}
\end{alignat}
Here $y_\textrm{min} = 0$ for massless partons and $y_\textrm{max}$ is
given by \eq(\ref{ymax}). The reduction of this measure for the simple case  $a_\ell\,=\,1$ has been presented in \cite{Chung:2010fx}. In this work, we have used the parametrization\footnote{We thank Z. Nagy and D. Soper for useful discussions concerning the parametrization of the integration measure.}
\begin{eqnarray*}
\lefteqn{d\xi_{p}\,=\,\frac{(2\,p_{\ell}\cdot Q)^{1-\vareps}}{16}\,\frac{\pi^{-\frac{5}{2}+\vareps}}{\Gamma\lb \frac{1}{2}-\vareps\rb}\times}\\
&&\,\int^{y_\text{max}}_{0}\,dy\,y^{-\vareps}\,\lambda^{1-2\,\vareps}\,\int^{1}_{0}\,dz\,\left[z\,(1-z)\right]^{-\vareps}\,\int^{1}_{0}\,dv\,\left[v\,(1-v)\right]^{-\frac{1+2\,\vareps}{2}}.
\end{eqnarray*}
In the center-of-mass system of $\ph_\ell,\,\ph_j$, where $\ph_k$ defines the $x-z$ plane, $z$ and $v$ parametrize the polar and azimuthal angles of $\ph_j$ respectively. 
\subsection{Generalized final state subtraction terms}\label{sec:subtr}
In this section we present
results for the subtraction terms ${\cal D}_{\ell}$ and their
integrated counterparts ${\cal V}_{\ell}$ for final state emitters, where $a_\ell\,\neq\,1$. Results for the simpler case of maximally two final state partons as well as initial state emitters have been presented in \cite{Chung:2010fx}. The integrated subtraction terms $\mathcal{V}_\ell$ contain integrals which depend on maximally two additional variables and need to be integrated numerically. 
In the expressions below, we leave out a common factor
$4\pi\al_{s}$ in the expressions for the squares $v_{\ell}^2$ of the
splitting amplitudes; the (integrated) subtraction
terms ($\mathcal{V}_{\ell}$) ${\cal D}_{\ell}$ contain all factors. We will summarize the scheme in Section \ref{sec:fin}. We used the Mathematica package HypExp \cite{Huber:2005yg,Huber:2007dx} in some of our calculations.

\subsubsection{Parameters}\label{sec:pars}

In this paper, we use the labeling
$\D_{f_{\ell}\hat{f}_{\ell}\hat{f}_{j}}$ and
$\mathcal{V}_{f_{\ell}\hat{f}_{\ell}\hat{f}_{j}}$ for a process with
the splitting $p_{\ell}\,\rightarrow\,\ph_{\ell}+\ph_{j}$. For
final state splittings, the subtraction terms can be
expressed through the variables
\begin{eqnarray}
&&y\,=\,\frac{\ph_{\ell}\cdot \ph_{j}}{p_{\ell}\cdot Q} \quad \mbox{and}\quad 
z\,=\,\frac{\ph_{j}\cdot n_{\ell}}{P_{\ell}\cdot n_{\ell}}\,,
\end{eqnarray}
with
\beq
P_{\ell}\,=\,\ph_{\ell}+\ph_{j},\;n_{\ell}\,=\,\frac{\gamma}{\lambda}\,Q-\frac{a_\ell}{\lambda}P,\;p_{\ell}\cdot Q\,=\,P_{\ell}\cdot Q-\ph_{\ell}\cdot\ph_{j}\,,
\eeq
where we additionally introduced
\begin{\eqn*}
\gamma\,=\,\frac{1+\lambda+y}{2},\,x_0\,=\,\frac{1-\lambda+y}{1+\lambda+y}.
\end{\eqn*}

\subsubsection{Collinear subtractions}\label{sec:finsq}
We first consider the collinear part of the subtraction terms, which are given by the first term in Eqn. (\ref{splitsplittingfunctions335}). These terms do not contain any soft or combined soft/ collinear singularities, \ie they only contain single poles $\sim\vareps^{-1}$ and do not depend on a specific spectator $k$.
\subsubsection*{qqg, $\bar{\text{q}}\bar{\text{q}}$g}
The squared splitting amplitude for final state $qqg$ couplings in the
case of massless quarks is given by
\begin{\eqn}\label{eq:qqgfinsq}
v_{qqg}^{2}\,=\,\frac{2}{y\,(p_\ell \cdot Q)}\,\left\{\left[ 1\,+\,\frac{(\lambda-1+y)^{2}+4\,y}{4\,\lambda}\right]\,F_\text{eik}\,+\,\frac{D-2}{4}\,z\,\left[ 1+y+\lambda\right]\right\},
\end{\eqn}
where
\begin{\eqn}\label{eq:feikfin}
F_\text{eik}\,=\,2\,\,\lb\,-1\,+\,\frac{1+x_{0}}{x_{0}+z\,(1-x_{0})}\,-\,\frac{x_{0}}{(x_{0}+z\,(1-x_{0}))^{2}}\rb.
\end{\eqn}
Thus we have 
\begin{eqnarray}\label{eq:dqqg_fin}
\D^{\text{coll}}_{qqg}&=&\frac{4\pi\al_{s}}{2}\,C_{F}\,\lb v_{qqg}^{2}-v_{\text{eik}}^{2} \rb\nonumber\\
&=&\frac{4\,\pi\,\al_s}{y\,(p_\ell \cdot Q)}\,C_F\,\left\{\frac{(\lambda-1+y)^{2}+4\,y}{4\,\lambda}\,F_\text{eik}\,+\,\frac{D-2}{4}\,z\,\left[ 1+y+\lambda\right]\right\},
\end{eqnarray}
and the integrated
subtraction term is 
\begin{eqnarray}\label{eq:vqqg_fin}
\lefteqn{\mathcal{V}_{qqg}^{\text{coll}}\,=\,
\frac{\al_{s}}{4\,\pi}\,C_{F}\,\frac{1}{\Gamma(1-\vareps)}\,\lb\frac{2\,\pi\,\mu^{2}}{p_{\ell}\cdot Q} \rb^{\vareps}\,\times\,
\Bigg\{-\frac{1}{\vareps}+\,4\,I_{3}(a_\ell)}\nonumber\\
&& +\frac{1}{2} \left[(9-7 a_\ell) (a_\ell-1) \log (a_\ell-1)+a_\ell (7 a_\ell-16) \log (a_\ell)-7 \log (\ymax)\right.\nonumber\\
&&\left.-a_\ell (2 \ymax+7) -7 \ymax-4\right] \Bigg\},
\end{eqnarray}
with
\begin{eqnarray}\label{eq:i3}
I_3(a_\ell)&=&-\int^{y_\text{max}}_{0}\,dy\,\left[\frac{(\lambda-1+y)^{2}}{4\,y}+1\right]\,\frac{(1+y)\ln\,x_{0}}{\lambda}.
\end{eqnarray}
\subsubsection*{gq$\bar{\text{q}}$, g$\bar{\text{q}}$q}
The $gq\bar{q}$ splitting function for massless quarks, keeping the gluon helicity for the mother parton, is given by
\begin{\eqn*}
\langle \nu |v^2_{gq\bar{q}}|\nu' \rangle \,=\,
\frac{1}{\ph_\ell\cdot \ph_j}\,\left[-g_{\nu\,\nu'}\,-\,2\frac{k_{\perp,\nu}\,k_{\perp,\nu'}}{\ph_\ell\cdot \ph_j} \right],
\end{\eqn*}
where $k_\perp$ can easily be obtained from a Sudakov parametrization as
\begin{eqnarray}\label{eq:kt}
k_\perp&=&\ph_\ell-\frac{1}{\lambda}\,\left[ P_\ell\,\gamma\,\lb 1-z\,(1+x_0) \rb+y\,Q\,(2\,z\,-1) \right],
\end{eqnarray}
with $k_\perp\cdot p_\ell\,=\,k_\perp\cdot n_\ell\,=\,0$. If there is no explicit helicity dependence in the Born-type matrix element, we have
\begin{\eqn}
v_{gq\bar{q}}^{2}\,=\,\frac{2}{y\,p_{\ell}\cdot\, Q}\,\lb 1-\vareps\,-2\,z\,(1-z) \rb.
\end{\eqn}
We obtain for the
subtraction terms
\begin{\eqn}\label{eq:dgqq_fin}
\langle \nu|\D_{gqq}|\nu'\rangle\,=\,{4\pi\al_{s}}\,T_{R}\langle \nu\,|v_{gq\bar{q}}^{2}|\nu'\rangle,\,\D^\text{av}_{gqq}\,=\,\frac{4\pi\al_{s}}{2\,(1-\vareps)}\,T_{R}\,v_{gq\bar{q}}^{2}.
\end{\eqn}
Integrating this over the unresolved one-parton phase space yields
\begin{eqnarray}\label{eq:vgqq_fin}
\mathcal{V}_{gqq}(a_\ell)\,=\,\frac{T_{R}}{\pi}\,\frac{\al_{s}}{\Gamma(1-\vareps)}\,\lb\frac{2\,\pi\,\mu^{2}}{p_{\ell}\cdot Q}\rb^{\vareps}\,\left[-\frac{1}{3\,\vareps}\,-\,\frac{8}{9}\,+\,\frac{1}{3}\,\left[(a_\ell-1)\,\ln(a_\ell-1)-a_\ell\,\ln\,a_\ell\right]\right].\nn\\
&&
\end{eqnarray}
\subsubsection*{ggg}
The total (unaveraged) splitting amplitude squared, in the helicity basis of the mother parton $p_\ell$, is given by
\begin{eqnarray}\label{eq:ggg_split}
\langle \nu|v^2_{ggg}|\nu'\rangle&=&\frac{1}{2\,(\ph_\ell\cdot \ph_j)^2}\,\left\{-g^{\nu\nu'}(\ph_\ell \cdot D_j \cdot \ph_\ell\,+\,\ph_j \cdot D_\ell \cdot \ph_j)+k_\perp^\nu\,k_\perp^{\nu'}\,\mbox{Tr}[D_\ell \cdot D_j]  \right\},\nn\\
&&
\end{eqnarray}
with
\begin{eqnarray*}
\ph_{\ell} \cdot D_{j} \cdot \ph_{\ell}&=&\frac{2\,y\,p_{\ell} \cdot Q}{x_{0}+z\,(1-x_{0})}\,\left[1\,-\,z\,(1-x_{0})\,-\,\frac{x_{0}}{x_{0}+z\,(1-x_{0})}
  \right],\\
\ph_{j} \cdot D_{\ell} \cdot \ph_{j}&=&\frac{2\,y\,p_{\ell} \cdot Q}{1-z\,(1-x_{0})}\,\left[x_{0}\,+\,z\,(1-x_{0})\,-\,\frac{x_{0}}{1-z\,(1-x_{0})}
  \right],\\
\mbox{Tr}\left[D_{\ell}\cdot D_{j}\right]&=&D-2-2\,\Delta+\Delta^{2}
\end{eqnarray*}
and
\begin{\eqn}\label{eq:delta}
\Delta\,=\,\frac{\hat{Q}^{2}\,(\ph_{\ell} \cdot \ph_{j})}{(\ph_{\ell}
  \cdot 
\hat{Q})\,(\ph_{j} \cdot \hat{Q})}\,=\,\frac{2\,x_{0}}{(x_{0}+z\,(1-x_{0}))\,(1-z\,(1-x_{0}))},
\end{\eqn}
and $k_\perp$ again given by Eqn. (\ref{eq:kt}); if the Born matrix element is helicity independent, we have
\begin{eqnarray*}
v_{ggg}^{2}\,=\,\frac{1}{2\,(\ph_{\ell}\cdot\ph_{j})^{2}}\left\{(D-2)\,\left[\ph_{\ell}\cdot D_{j}\cdot \ph_{\ell}+\ph_{j}\cdot D_{\ell} \cdot \ph_{j}\right]\,-\,k_{\perp}^{2}\mbox{Tr}\left[D_{\ell} \cdot D_{j}
  \right] \right\},
\end{eqnarray*}
with
\begin{\eqn}\label{eq:ktsq}
k_{\perp}^{2}\,=\,-2\,y\,z\,(1-z)\,p_{\ell} \cdot Q.
\end{\eqn}
Instead of using this as a subtraction term, however, we proceed in a
different way, and define a subtraction term that only contains soft singularities from particle $j$\cite{Nagy:2008ns}: 
We introduce
\begin{\eqn}
\langle \nu|v_{ggg,\text{sub}}^{2}|\nu'\rangle\,=\,\langle \nu|v_{2}^{2}-v_{3}^{2}|\nu'\rangle\,=\,-\frac{g^{\nu\nu'}}{2\,(\ph_{\ell}
  \cdot 
\ph_{j})^{2}}\,\left[\ph_{\ell}\cdot D_{j}\cdot \ph_{\ell}-\ph_{j}\cdot D_{\ell}\cdot\ph_{j} \right],
\end{\eqn}
where $v_{2,3}$ are defined corresponding to \eqs (2.40)-(2.42) in
\cite{Nagy:2008ns}. This leads to
\begin{eqnarray*}
\langle \nu|\tilde{v}_{ggg}^{2}|\nu'\rangle &=&\langle \nu|v^{2}_{ggg}+v_{ggg,\text{sub}}^{2}|\nu'\rangle\\
&=&\frac{1}{2\,(\ph_{\ell}
  \cdot \ph_{j})^{2}}\left\{-2\,g^{\nu\nu'}\,\ph_{\ell}\cdot D_{j}\cdot
  \ph_{\ell}\,
+\,k_{\perp}^{\nu}\,k_\perp^{\nu'}\mbox{Tr}\left[D_{\ell}\cdot D_{j}
  \right] \right\}\,,
\end{eqnarray*}
which is the subtraction term for each gluon emission.
The first part is the unaveraged eikonal splitting function; if we
combine this with the interference term, we have
\begin{\eqn}
\langle \nu|\tilde{v}_{ggg}^{2}-v^{2}_\text{eik}|\nu'\rangle\,=\,\frac{k_{\perp}^{\nu}\,k_\perp^{\nu'}}{2\,(\ph_{\ell}
  \cdot \ph_{j})^{2}}\,\left[D-2-\Delta(2-\Delta)\right].
\end{\eqn}
The collinear subtraction term reads
\begin{\eqn}\label{eq:dggg_fin}
\langle \nu|D^{\text{coll}}_{ggg}|\nu'\rangle\,=\,\frac{2\,\pi\,\al_{s}}{1-\vareps}\,C_{A}\, \frac{k_{\perp}^{\nu}\,k_\perp^{\nu'}}{(\ph_{\ell}
  \cdot \ph_{j})^{2}}\,\left[D-2-\Delta(2-\Delta)\right].
\end{\eqn}
If there is no angular correlation in the Born-type matrix element, we can replace $k_\perp^\nu\,k_\perp^{\nu'}\,\rightarrow\,-k_\perp^2$ in the above expressions, and equally need to multiply by  ${1}/{2(1-\vareps)}$.\\

Note that the above reshuffling of singular terms requires that for a final state with $g(\ph_1)g(\ph_2)$, both combinations $(i,j)\,=\,(1,2),(2,1)$ need to be taken into account; the factor $\frac{1}{2}$ which is included in Eqn. (\ref{eq:ggg_split}) and all subsequent expressions guarantees a correct mapping of the singularity structure.\\
  
Integrating and taking all averaging factors into account gives
\begin{eqnarray}\label{eq:vggg_fin}
\lefteqn{\mathcal{V}_{ggg}^{\text{coll}}\,=\,\mu^{2\vareps}\,\frac{4\,\pi\,\al_{s}}{2\,(1-\vareps)}\,C_{A}
\,\int\,d\xi_{p}\,\lb\tilde{v}_{ggg}^{2}-v^{2}_\text{eik}\right)}\nn\\
&=&\lb\frac{2\,\pi\,\mu^{2}}{p_{\ell}\cdot \hat{Q}}\rb^{\vareps}\,\frac{1}{\Gamma(1-\vareps)}\,\frac{\al_{s}}{2\pi}\,C_{A}
\left[-\frac{1}{6\,\vareps}\,-\,\frac{4}{9}\,+\,\frac{1}{6}\,\left[(a_\ell-1)\,\ln(a_\ell-1)\,-\,a_\ell\ln\, a_\ell\right]\,+\,I_\text{fin}(a_\ell)\right]\nn\\
&&
\end{eqnarray}
with
\begin{eqnarray}\label{eq:ifin}
I_\text{fin}(a_\ell)
&\stackrel{a_\ell\neq\,1}{=}&a_\ell\,\left\{1-\sqrt{a_\ell}\,\ln\lb\frac{\sqrt{a_\ell}+1}{\sqrt{a_\ell-1}}\rb-\ln\lb\frac{a_\ell}{a_\ell-1}\rb\right.\nn\\
&&\left.\,+\,8\,a_\ell\int^{\ymax}_{0}\,dy\,\frac{y\,\ln\,x_{0}}{\lambda^{2}\,(1+y)^{3}}\,\left[a_\ell\,y-(1+y)^{2}\right] \right\}\nn\\
&\stackrel{a_\ell=1}{=}&-\frac{3}{8}\pi^{2}\,+\,\frac{7}{2}.
\end{eqnarray}

\subsubsection{Soft and soft/collinear subtractions}\label{sec:soft}
We now  discuss the integration of the interference term, which is
given by the second contribution in
Eqn. (\ref{splitsplittingfunctions335}). This term does not depend on
the specific nature of the splitting, \ie it is universal; it contains
all soft and soft/ collinear singularities and equally depends on a
spectator parton $k$. Parton $j$ needs to be a gluon, otherwise this contribution vanishes.\\

We start from the definition of the interference term in Eqn. (\ref{interferencespinaveragedsplittingfunction}): 
\begin{\eqn*}
\frac{1}{4\,\pi\,\al_{s}}\Delta\,W_{\ell k}\,=\,\frac{2\,(\ph_{\ell} \cdot
  \ph_{k})\,(\ph_{\ell} 
\cdot \hat{Q})}{(\ph_{\ell} \cdot \ph_{j})\,\lb(\ph_{j} \cdot
\ph_{k})\,(\ph_{\ell} 
\cdot \hat{Q})+(\ph_{\ell} \cdot \ph_{j})(\ph_{k} \cdot \hat{Q})\rb}
\end{\eqn*}
and the subtraction term
\begin{\eqn}\label{eq:Dif_fin}
\D^{\text{if}}(\ph_\ell,\ph_j,\ph_k)\,=\,C_{\ell k}\,\Delta W_{\ell k},
\end{\eqn}
where $C_{\ell k}$ is given by Eqn. (\ref{eq:clk}). Note
that the above expression holds also for cases where the mother parton
is a gluon, as the interference terms are diagonal in helicity
space\footnote{That is, for helicity dependent Born-type matrix elements $\M$, where a spin correlation tensor $\mathcal{T}^{\mu\nu}$ is defined such that $-g_{\mu\nu}\mathcal{T}^{\mu\nu}\,=\,|\M|^2$,  the interference subtraction term is given by 
\begin{\eqn*}
-g_{\mu\nu }D^\text{if}\mathcal{T}^{\mu\nu}\,=\,D^\text{if}|\M|^2
.
\end{\eqn*}}.

We obtain for the integrated subtraction term
\begin{eqnarray}\label{eq:vif}
\mathcal{V}_{\ell k}^{\text{if}}&=&\mu^{2\,\vareps}\,C_{\ell k}\,\int\,d\xi_{p}\,(\Delta\,W_{\ell k})\,=\,
\lb\frac{2\,\mu^{2}\,\pi}{p_{\ell}\cdot Q}\rb^{\vareps}\frac{\al_{s}}{\pi}\,\frac{1}{\Gamma(1-\vareps)}C_{\ell k}\nn\\
&&\times \left\{\frac{1}{2\,\vareps^{2}}\,+\,\frac{1}{\vareps}\,\left[1\,+\,\frac{1}{2}\,\ln\lb \tilde{a}^{(\ell k)}_{0}+a_\ell\rb\right]\,\right.\left. -\,\frac{\pi^{2}}{6}\,+\,3\,-\,2\,\ln\,2\,\ln\lb \tilde{a}^{(\ell k)}_{0}+a_\ell \rb\right. \nn \\
&&\left.\,+\,\frac{1}{\pi}\left[I^{(b)}_\text{fin}\lb\frac{\tilde{a}^{(\ell k)}_{0}}{a_\ell}\rb\,+\,I^{(d)}_\text{fin}(a_\ell,\tilde{a}^{(\ell k)})\,+\,I^{(e)}_\text{fin}(a_\ell)\right]\right.\nn\\
&&\left. +\,\ln\,a_\ell\,\left[2\,\ln\,2\,-\,\frac{1}{4}\,\ln\,a_\ell\,+\,\frac{1}{2}\,\ln\,\lb \tilde{a}^{(\ell k)}_{0}\,+\,a_\ell\rb \,+\,1\right]\,\right\},\nn\\
&&
\end{eqnarray}
with
\begin{eqnarray}\label{eq:ibfin}
\lefteqn{I^{(b)}_\text{fin}(b) \,=\, \frac{\pi}{2}\Bigg[\int^{1}_{0}\,\frac{du}{u}\,\Bigg\{2\ln2\,
+\,\frac{1}{\sqrt{1+4\,b(1+b)\,u^{2}}}\left.\times\,\ln\left[\frac{(1-u)}{\lb 1+2\,b\,u\,+\,\sqrt{1+4\,b\,(1+b)\,u^{2}}
         \rb^{2}}
    \right]\right\}}\nonumber\\
&&\hspace*{5mm}+\,2\,\ln\,2\,\ln\lb 1+b\rb\,+\,\frac{1}{2}\,\ln^{2}\,\lb
    1+b\rb\,+\,\frac{5}{2}\text{Li}_{2}\,\lb\frac{b}{b+1}\rb -\,\frac{1}{2}\text{Li}_{2}\,\left[\lb\frac{b}{b+1}\rb^{2}\right]\Bigg],\nn\\
\lefteqn{I^{(d)}_\text{fin}(a_\ell,\tilde{a}^{(\ell,k)})
\,=\,\pi\int^{1}_{0}\,\frac{du}{u}\,\int^{1}_{0}\,\frac{dx}{x}}\nonumber\\
&&\times\,\Bigg\{ \gamma_\ell\,x\left[\frac{ 1-x+x_{0,\ell}\,\left[\lambda_\ell\,\frac{\tilde{a}^{(\ell,k)}}{a_\ell}\,+\,2 \right]}{\sqrt{\left[ A^{(\ell k)}\,(1+x_{0,\ell}-x)+x_{0,\ell}\lb \lambda_\ell\frac{\tilde{a}^{(\ell,k)}}{a_\ell}+1 \rb \right]^2-[B^{(\ell k)}]^2(1+x_{0,\ell}-x)^2}}-1 \right]\nn\\
&&+\,x\,-\,\frac{1}{\sqrt{1+4\,\frac{u^2\,\tilde{a}^{(\ell,k)}_0}{a_\ell}\left[1+\frac{\tilde{a}^{(\ell,k)}_0}{a_\ell}\right]}}\Bigg\}\nn\\
\lefteqn{I^{(e)}_\text{fin}(a_\ell)\,=\,\pi\,\int^{1}_{0}\,dx\,\frac{1-x}{x}\,\ln\left[\frac{\delta_\ell\,a_\ell}{x}\right].}
\end{eqnarray}
We have introduced 
\begin{eqnarray*}
&&A^{(\ell k)}(p_\ell,p_k)\,=\,z_\ell\,\gamma_\ell\,\frac{p_\ell\cdot\ph_k}{\ph_k\cdot P_\ell}\,+\,(1-z_\ell)\,\tilde{z}^{(\ell k)},\;
B^{(\ell k)}(p_\ell,p_k)\,=\,2\,\sqrt{z_\ell\,\tilde{z}^{(\ell k)}\,(1-z_\ell)\,(1-\tilde{z}^{(\ell k)})},
\end{eqnarray*}
and 
\begin{eqnarray*}
z_\ell\,=\,\frac{x-x_{0,\ell}}{1-x_{0,\ell}},&&\tilde{z}^{(\ell k)}\,=\,\frac{y_\ell}{\gamma_\ell}\tilde{a}^{(\ell k)},\\
\tilde{a}^{(\ell k)}\,=\,\frac{\ph_{k}
  \cdot n_{\ell}}{\ph_{k} \cdot P_\ell},&&\tilde{a}^{(\ell k)}_{0}\,=\,
\tilde{a}^{(\ell k)}(y_\ell=0)\,=\,\frac{p_{k} \cdot n_{\ell}}{p_{k} \cdot p_\ell}\,\\
\delta_\ell(x)&=&\frac{a_\ell\,-\,\sqrt{a_\ell^{2}-a_\ell\,\frac{4\,x}{(1+x)^{2}}}}{2\,x}\,(1+x)^{2}\,-\,1,
\end{eqnarray*}
where $y_\ell$ in all above expressions is defined by\footnote{We thank Z. Nagy for providing us with this variable transformation for the interference terms.}
\begin{\eqn*}
y_\ell\,:=\,\delta_\ell(x)\,u.
\end{\eqn*}
We here made the dependence on the momenta $p_\ell,p_k$ explicit in the labeling of the variables $\lambda,\,\gamma,\,x_0,\,...$, which are all defined according to Sections \ref{sec:mommap} and \ref{sec:pars} respectively.\\

In terms of the Born-type kinematics, $P_\ell$ can easily be recovered from Eqn.~(\ref{totalmomentumfinalstatepellpj}); $\ph_k$ needs to be reconstructed in the Born-type integrations according to
\begin{\eqn}
\ph_k\,=\,\Lambda (\widehat{K},K)\,p_k,
\end{\eqn}
with the Lorentz transformation defined according to Eqn. (\ref{eq:LTini}) and with
\begin{\eqn*}
K\,=\,Q-p_\ell,\;\widehat{K}\,=\,Q\lb 1-\frac{\gamma_\ell\,x_{0,\ell}}{a_\ell} \rb-\lambda_\ell\,p_\ell.
\end{\eqn*}
Finally, note that
\begin{\eqn*}
a_\ell\,+\,\tilde{a}_0^{(\ell,k)}\,=\,\frac{p_k\cdot Q}{p_k\cdot p_\ell}.
\end{\eqn*}
\subsection{Final expressions}\label{sec:fin}
In this section, we describe how the expressions in the last subsections should be combined to provide the subtraction terms $d\sigma_A$ and their integrated counterparts $\int_1 d\sigma_A$.\\

\noindent
The complete parton level contribution is given by the sum of $\sigma^{\text{LO}}_{ab}$ and $\sigma^{\text{NLO}}_{ab}$, with
\begin{alignat}{53}
\sigma^{\text{LO}}_{ab}&= \int_m d\sigma^B_{ab}(p_a,p_b),\notag \\
\sigma^{\text{NLO}}_{ab}&=\int_{m+1}d\sigma^R_{ab}(\ph_a,\ph_b)+
\int_{m}d\sigma^V_{ab}(p_a,p_b)+\int_{m}d\sigma^C_{ab}(p_a,p_b,\mu_F^2).
\nn
\end{alignat}
The NLO contribution can be split into
\begin{alignat}{53}
\label{eq:sig_nlo}
\sigma_{ab}^{\text{NLO}}&=  \int_{m+1} \left[ d\sigma_{ab}^{R}(\ph_a,\ph_b) -
d\sigma_{ab}^{A}(\ph_a,\ph_b) \right]                    \notag \\
&+ 
\int_{m}\, \left[\int d\sigma^{V}_{ab}(p_a,p_b)+ \int_1d\sigma^{A}_{ab}(\ph_a,\ph_b)+d\sigma_{ab}^{C}(p_a,p_b,\mu_F^2)\right]_{\vareps=0},\nn
\end{alignat}
where $\int_1 d\sigma^A_{ab}+ d\sigma^C_{ab}$ can be written as
\begin{alignat}{53}
&\int_{m}  \left[ \int_{1} d\sigma^A_{ab}(\ph_a,\ph_b)+ d\sigma^C_{ab}(p_a,p_b,\mu_F^2) \right]    \notag \\
=&
\int_m  d\sigma_{ab}^{B}(p_a,p_b) \otimes { I}(\vareps)
+ \int_0^1 dx \int_m  d\sigma_{ab}^{B}(x\ph_a,p_b)\otimes
\left[ { K}^a(x\,\ph_a) +   { P}(x,\mu_F^2)  \right]                             \notag \\
+&  \int_0^1 dx \int_m d\sigma_{ab}^{B}(\ph_a,x\ph_b)\otimes
\left[ { K}^b(x\ph_b) + { P}(x,\mu_F^2)  \right], 
\nn
\end{alignat}
where the insertion terms $K,\,P$ only appear in the case of initial state partons, \cf Appendix \ref{app:cts}.
All observables, as well as infrared safety of the Born level
contributions, need to be introduced in terms of jet functions as
discussed in Section \ref{Observableindependentformulation}, \cf Eqn. (\ref{eq:jetincl}). 
For an incoming lepton, the
collinear counterterm is set to zero and the PDF is replaced by a
structure function
$f_{i/I}^{\text{ew}}\,=\,\delta(1-\eta_{i})$.

In the following, we discuss the specific form of $d\sigma^{A}_{ab}(p_a,p_b)$ which corresponds to the subtraction term in the real emission contribution of the process, as well as the integrated $D$-dimensional counterterm $\int_1d\sigma^{A}_{ab}(p_a,p_b)$. In general, the subtraction term can be split into contributions originating from all possible emitters $\ph_\ell$ \footnote{In the following, we omit the jet functions for notational reasons; however, full expressions should always be read according to \eq (\ref{eq:jetincl}) where all jet functions are included.}:
\begin{\eqn}\label{eq:master_sub}
d\sigma^{A}_{ab}(\ph_a,\ph_b)\,=\,\sum_{\ell} d\sigma^{A,\ell}_{ab}(\ph_a,\ph_b),
\end{\eqn}
where $\ph_\ell$ can denote an initial or final state particle. We have for each contribution 
\begin{eqnarray}\label{eq:counter_ini}
d\sigma^{A,\ell}_{ab}(\hat{p}_a,\hat{p}_b)&=&\frac{N_{m+1}}{\Phi_{m+1}} \int_{m+1} 
\sum_{j\,\neq\,\ell} \D_{f_\ell\,\hat{f}_\ell\,\hat{f}_j}(\ph_\ell,\ph_j)\,\otimes\,|\M(p)|^2_{m;f_\ell}, 
\end{eqnarray}
where $|\M|^2_{m;f_\ell} $ denotes the squared Born matrix element with a flavour $f_\ell$ of the mother parton; the extension for cases where there is an angular dependence of the Born-type matrix element is straightforward. The momenta $\left\{ p_m\right\}$ are determined from $\left\{ \ph_m\right\}$ through the respective mapping. $N_{m+1}$ incorporates all symmetry factors of the $m+1$ process and $\Phi_{m+1}\,=\,2\,\hat{s}$ is the respective flux factor. For splittings where the mother parton is a gluon, we use the following conventions: for $g\,\rightarrow\,q\,\bar{q}$ final state splittings, we always choose $(\hat{f}_\ell,\hat{f}_j)\,=\,(q,\bar{q})$; for $g\rightarrow gg$, \ie a final state that contains $g(\ph_1) g(\ph_2)$, we need to consider {\sl both} combinations $(\ph_\ell,\ph_j)\,=\,(\ph_1,\ph_2),(\ph_2,\ph_1)$; we compensate this by introducing an additional factor $\frac{1}{2}$ in the respective (integrated) subtraction terms. This factor has already been accounted for in all expressions in Section \ref{sec:subtr}.\\

 The subtraction terms can be split into collinear and interference terms:
 \begin{\eqn}\label{eq:delljtot}
\D_{f_\ell \hat{f}_\ell \hat{f}_{j}}(\ph_\ell, \ph_j)\,=\,\D^{\text{coll}}_{f_\ell \hat{f}_\ell \hat{f}_{j}}(\ph_\ell, \ph_j)\,+\,\delta_{\hat{f}_{j},g}\sum_{k\,\neq\,(\ell,j)}\D^{\text{if}}(\ph_\ell,\ph_j,\ph_{k}),
\end{\eqn}
where $\D^{\text{if}}(\ph_\ell,\ph_j,\ph_{k})$ now denotes an interference contribution where $\ph_{k}$ acts as a spectator as discussed in Section \ref{Softsplittingfunction2242010}. Note that there is a {\sl unique momentum mapping} for each combination $(\ph_\ell,\ph_j)$  which is the same for all interference terms appearing in $\D_{f_\ell \hat{f}_\ell \hat{f}_{j}}(\ph_\ell, \ph_j)$.\\

\noindent
The integrated counterterms are given by the integrated form of \eq (\ref{eq:master_sub}):
\begin{\eqn*}
\int_{1}d\sigma^{A}_{ab}(\ph_a,\ph_b)\,=\,\int_{1}
\sum_{\ell} d\sigma^{A,\ell}_{ab}(\ph_a,\ph_b).
\end{\eqn*}
The collection of the integrated counterterms is then
straightforward: for each dipole that has been subtracted in the real
emission part, the respective integrated contribution to $I,\,K,\,P$
needs to be added to the virtual contribution as in \eq (\ref{eq:sig_nlo}).
Finally, our expressions have been derived on a matrix element level:
\begin{\eqn*}
\int_{1}\,|\M|^{2}_{m+1}\,\rightarrow\,\int_{1} \D \otimes\,|\M|^{2}_{m}\,=\,\mathcal{V}\otimes|\M|^{2}_{m};
\end{\eqn*}
on cross section level, we additionally have to take the flux as well as combinatorial factors into account\footnote{Correct counting of symmetry factors needs to be done explicitly in this expression; if all splitting multiplicities and symmetry factors are taken into account, we obtain a generic combinatoric factor $\frac{1}{2}$ for $ggg$ splittings, \cf Section 7.2 in \cite{Catani:1996vz}.}
\begin{eqnarray*}
\int_{1}d\sigma^{A}_{m+1;ab}(\ph_a,\ph_b)&=&\frac{N_{m+1}}{2\hat{s}}\,\int_{1}\D\,\otimes\,|\M|^{2}_{m}\,=\,\frac{N_{m+1}}{2\hat{s}}\mathcal{V}\,\otimes\,|\M|^{2}_{m},\\
\int_{m}\int_{1}d\sigma^{A}_{m+1;ab}&=&N_{m+1}\,\int_{m}\frac{1}{2\hat{s}}\mathcal{V}\,\otimes\,|\M|^{2}_{m}\,=\,\frac{N_{m+1}}{N_{m}}\,x_s\,\mathcal{V}\otimes\,\int_{m}d\sigma_{m},
\end{eqnarray*}
where the factors $N_{m},\,N_{m+1}$ account for possible symmetry
factors of the specific process, and where here $x_s\,=\,{s}/\hat{s}$  is the ratio of the partonic center-of-mass energies before and after the splitting; $x_s\,=\,1$ for final state emitters. We then obtain the relation
\begin{\eqn}\label{eq:v_ikp}
\sum\,\mathcal{V}\,=\,\frac{1}{x_s}\,\frac{N_{m}}{N_{m+1}}\lb
I\,+\,K\,+\,P \rb
\end{\eqn}
between the integrated splitting functions $\mathcal{V}$ given in
the next sections and the insertion operators $I,K,P$.

\section{Example: $e^+e^- \,\to\,$ 3 jets}\label{sec:appl}

\label{Threejetproduction2011}
                     
In this section we consider the simplest nontrivial process with more than two partons in the final state: three-jet production in
$e^+e^-$ annihilation. The next-to-leading order contributions to this process are well known \cite{Ellis:1980wv,Kuijf:1991kn,Giele:1993dj,Catani:1996jh}. We compare the results obtained from the implementation of our scheme and from a private implementation of the Catani Seymour scheme as well as \cite{Catani:1996jh}\footnote{We thank M. Seymour for help with the original code available from \cite{event2f}.}. We find complete agreement for the differential $C$ parameter \cite{Ellis:1980wv}, with integration errors on the percent level.\\

The leading order process we consider is given by  
\begin{\eqn}\label{eq:LO}
e^+\,e^-\,\longrightarrow\,q(p_1)\,\bar{q}(p_2)\,g(p_3).
\end{\eqn}
At next-to-leading order, two different real-radiation subprocesses contribute:
\bea
\label{eenlorealemission2011}
(A)\qquad e^+\, e^- &\to& \gamma^*(Q) \,\to\, q(\hat p_1)\, \bar q(\hat p_2)\,g(\hat p_3)\,      g(\hat p_4),\nn\\
(B)\qquad e^+\, e^- &\to& \gamma^*(Q) \,\to\, q(\hat p_1)\, \bar q(\hat p_2)\,q(\hat p_3)\,\bar q(\hat p_4). 
\eea 
For a complete next-to-leading order calculation, the virtual corrections need to be added to the leading order contribution.
In the following, we use the notation
\begin{\eqn*} 
x_i=2\,p_i\cdot Q/Q^2,\,
y_{ij}={s_{ij}}/{Q^2},\, s_{ijk}=s_{ij}+s_{ik}+s_{jk},
\end{\eqn*}
with $s_{ij}\,=\,(p_i+p_j)^2$.
Energy-momentum conservation leads to
\begin{\eqn*}
\sum_{i,j>i}y_{ij}\,=\,1,\;\sum_i\,x_i\,=\,2.
\end{\eqn*}
We equally follow the notation for matrix elements in Section \ref{sec:splitf}:
\begin{eqnarray*}
\left\langle\, \left\{p_i\right\} \,\right|\left.\, \left\{p_i\right\} \,\right\rangle &=&
\left\langle\, {\cal M}_3\lb  \left\{p_i\right\}  \rb \,\right|\left.\, {\cal M}_3\lb  \left\{p_i\right\}  \rb \,\right\rangle\,\equiv\,\left|{\cal M}_3\lb  \left\{p_i\right\}  \rb\right|^2,\\
\left\langle\, \left\{\ph_i\right\} \,\right|\left.\, \left\{\ph_i\right\} \,\right\rangle &=&
\left\langle\, {\cal M}_4\lb  \left\{\ph_i\right\}  \rb \,\right|\left.\, {\cal M}_4\lb  \left\{\ph_i\right\}  \rb \,\right\rangle\,\equiv\,\left|{\cal M}_4\lb  \left\{\ph_i\right\}  \rb\right|^2
\end{eqnarray*}
The total next-to-leading order contribution for process (\ref{eq:LO}) is then given by:
\begin{eqnarray}
\label{finalstructurenlo532011}
\lefteqn{\sigma^{NLO}_J= \int dPS_4\left[  \left|{\cal M}_4\lb\left\{\hat p_i\right\}\rb\right|^2 F_J^{(4)}\lb\left\{\hat p_i\right\}\rb - 
\sum_{\ell,j} \left\langle  q,r,s  \right|  {\cal D}_{\ell j}\lb\left\{\hat p_i\right\}\rb \left| q,r,s \right\rangle    F_J^{(3)} (q,r,s)
 \right]_A }\nn \\
&&+ \int dPS_4\sum_{\textrm{flavours}}
\left[  \left|{\cal M}_4\lb\left\{\hat p_i\right\}\rb \right|^2 F_J^{(4)}\lb\left\{\hat p_i\right\}\rb - 
\sum_{\ell,j} \left\langle  q,r,s  \right|  {\cal D}_{\ell j}\lb\left\{\hat p_i\right\}\rb  \left| q,r,s \right\rangle    F_J^{(3)} (q,r,s)
 \right]_B
 \nn \\
&&+ \int dPS_3   \Bigg\{ \left|{\cal M}_V \lb\left\{p_i\right\}\rb \right|^2 + 
\frac{1}{2}\left[\sum_{\ell,j} \mu^{2 \vareps}\int d\xi_p 
\left\langle  1,2,3  \right|  {\cal D}_{\ell j}\lb\left\{\ph_i\right\}\rb  \left| 1,2,3 \right\rangle \right]_A \Bigg.\nn \\
&&\Bigg.   +  \sum_{\textrm{flavours}}  \frac{1}{4}
\left[\sum_{\ell,j} \mu^{2 \vareps}\int d\xi_p 
\left\langle  1,2,3  \right|  {\cal D}_{\ell j}\lb\left\{\ph_i\right\}\rb   \left| 1,2,3 \right\rangle \right]_B
 \Bigg\} F_J^{(3)}  \lb \left\{ p_i \right\} \rb
\end{eqnarray}
where $\ell,j$ sum over all possible pairs in the real emission phase space, $\ph_i$ and $p_i$ denote momenta belonging to Born and real emission kinematics, and where $q,r,s \,\in\, \{1,2,3\}$ denote the mapped momenta in the real emission subtraction terms for the Born-type matrix elements. The factors $\frac{1}{2},\,\frac{1}{4}$ in the integrated subtraction terms correspond to the process-dependent symmetry factors in the real emission contributions. In the symbolic notation above, $\int\,dPS_n$ contains all symmetry and flux factors for the respective phase space\footnote{Note that  $\int\,dPS_3$ and $\int\,dPS_4$ are defined slightly differently in \cite{Catani:1996jh,Catani:1996vz}.}.
\subsection{Tree level result}
In the following, we follow the procedure of \cite{Ellis:1980wv}, \ie we normalize our observables according to
\begin{\eqn*}
\frac{1}{\sigma_{0}}\,d\sigma^\text{NLO}_{J},
\end{\eqn*}
where $\sigma_0$ denotes the total cross section for the process
\begin{\eqn}\label{eq:eeqq}
e^+\,e^-\,\longrightarrow\,q\,\bar{q}
\end{\eqn}
given by \cite{Ellis:1980wv}
\begin{\eqn*}
\sigma_{0}\,=\,\frac{4\,\pi\,\al^2}{3\,s}\,C_A\,q_f^2
\end{\eqn*}
for a quark with flavour $f$ and charge $q_f$.
We normalize the matrix elements according to
\begin{\eqn*}
\sigma^{(n)}_J\,=\,\frac{1}{2\,s}\,S\,\int\,d\Gamma_n\,|\M_n|^2\,F_J^{(n)},
\end{\eqn*}
where $d\Gamma_n\,=\,\prod_i\,\left[\frac{d^4p_i}{(2\pi)^4}\,\delta\lb p_i^2-m_i^2 \rb\right]\,\delta^{(4)}\lb \sum_\text{in}p_\text{in}-\sum_i\,p_i \rb$, $s$ denotes the center-of-mass energy and
\begin{eqnarray*}
S
    &=&   \left\{  \begin{array}{cl}
          {1}/{2!} &\quad \textrm{for} \quad  \gamma^*\to q\,\bar q\,g\,g   \\
          & \\
          {1}/{2!}\times {1}/{2!} &\quad \textrm{for}\quad  \gamma^*\to q\,\bar q\,q\,\bar q    \\
                 \end{array}\right.
\end{eqnarray*}
is the symmetry factor of the respective process. We obtain the well-known relations between the matrix elements of the processes (\ref{eq:eeqq}) and (\ref{eq:LO})  \cite{Ellis:1980wv}:
\begin{\eqn*}
\frac{1}{4}|\M_3|^2\,=\,\frac{8\,\pi\,\al_s}{s}\,C_F\,\frac{x_1^2+x_2^2}{(1-x_1)\,(1-x_2)}\,\frac{1}{4}\langle|\M_2|^2 \rangle
\end{\eqn*}
where $|\M_2|^2$ has been averaged over the emission angles, as well as
\begin{\eqn*}
\sigma_{3,J}\,=\,\frac{\al_s}{2\,\pi} C_F \int dx_1 dx_2 \Pi_i \left[ \Theta(1-x_i) \Theta(x_i)\right]\Theta(x_1+x_2-1) \frac{x_1^2+x_2^2}{(1-x_1) (1-x_2)} \sigma_0 F^{(3)}_J(x_1,x_2)
\end{\eqn*}
for jet observables.
The gluon-helicity dependent squared matrix element for the leading order process (\ref{eq:LO}) 
is given by \cite{Catani:1996jh}
\beq
{\cal T}^{\mu \nu}(p_1,p_2,p_3) = - \frac{1}{x_1^2 + x_2^2}\,
 \left|{\cal M}_3(p_1,p_2,p_3)\right|^2 \,T^{\mu \nu}, 
\eeq
where
\begin{alignat}{5}
\label{}
  T^{\mu\nu} &=
    +2\,                            \frac{p_1^\mu p_2^\nu}{Q^2}
    +2\,                            \frac{p_2^\mu p_1^\nu}{Q^2}
    -2\,\frac{1-x_1}{1-x_2}   \,    \frac{p_1^\mu p_1^\nu}{Q^2}
    -2\,\frac{1-x_2}{1-x_1}   \,    \frac{p_2^\mu p_2^\nu}{Q^2}      \notag \\
    &
    -\frac{1-x_1-x_2+x_2^2}{1-x_2}\,\left[
                                    \frac{p_1^\mu p_3^\nu}{Q^2}
    +                               \frac{p_3^\mu p_1^\nu}{Q^2}
                                    \right]
    -\frac{1-x_2-x_1+x_1^2}{1-x_1}\,\left[
                                    \frac{p_2^\mu p_3^\nu}{Q^2}
    +                               \frac{p_3^\mu p_2^\nu}{Q^2}
                                    \right]                          \notag \\
   &
    +\left(1+\frac 1 2\, x_1^2+\frac 1 2\, x_2^2-x_1-x_2\right)\,g^{\mu\nu}. 
\end{alignat}
${\cal T}_{\mu \nu}$ obeys
$g^{\mu \nu} {\cal T}_{\mu \nu} =- |{\cal M}_3|^2$. 

\subsubsection*{The virtual one-loop matrix element}

The virtual contribution to the process (\ref{eq:LO}) in
the $\msbar$ renormalization scheme is \cite{Ellis:1980wv,Catani:1996vz}
\begin{eqnarray}
\label{loopthreejets31}
\lefteqn{\left|{\cal M}_V(p_1,p_2,p_3) \right|^2\, =\,
\left|{\cal M}_3(p_1,p_2,p_3)\right|^2 \, \frac{\as}{2\pi}\, 
\left(\frac{4\pi\mu^2}{Q^2}\right)^\vareps\,
\frac{1}{\Gamma(1-\vareps)}}
\nn \\
&\times&
\left\{ \,-\frac{1}{\vareps^2} \left[ \left(2C_F -C_A\right) y_{12}^{-\vareps} +
C_A \left(y_{13}^{-\vareps} + y_{23}^{-\vareps} \right) \right]
-\frac{1}{\vareps}\, \left( 3\,C_F + \frac{11}{6}C_A -\frac{2}{3}n_f\,T_R \right)
\right. \nn \\
&&+ \left. \frac{\pi^2}{2}\,(2\,C_F+C_A) - 8\, C_F \right\}
+ \frac{\as}{2\pi}\,\left[ F(y_{12},y_{13},y_{23}) +{\cal O}(\vareps) \right], 
\end{eqnarray}
where $F(y_{12},y_{13},y_{23})$ is defined in \eq (2.21) of \cite{Ellis:1980wv}.
\subsection{Subtraction terms}
In general, the subtraction terms are given by \eq (\ref{eq:delljtot}) as
 \begin{\eqn*}
\D_{\ell j}\,\equiv\,\D_{f_\ell \hat{f}_\ell \hat{f}_{j}}(\ph_\ell, \ph_j)\,=\,\D^{\text{coll}}_{f_\ell \hat{f}_\ell \hat{f}_{j}}(\ph_\ell, \ph_j)\,+\,\delta_{\hat{f}_{j},g}\sum_{k\,\neq\,(\ell,j)}\D^{\text{if}}(\ph_\ell,\ph_j,\ph_{k}),
\end{\eqn*}
where the first part is the collinear subtraction term, and the second the sum over all interference terms. For the processes considered here, the interference terms only contribute to $(A)$, while $(B)$ only contains collinear divergences. 
In the following sections we will construct the subtraction terms for subprocesses $(A)$ and $(B)$, respectively.

\subsubsection{Momentum mapping}

For all subtraction terms, the $m$-parton phase space momenta are mapped as described in section \ref{sec:mommapsub}; \ie, for a splitting
\begin{\eqn*}
p_\ell\,\rightarrow\,\ph_\ell+\ph_j,
\end{\eqn*}
we have
\begin{eqnarray*}
p_\ell&=&\frac{1}{\lambda}\,P_\ell-\frac{1 - \lambda + y}{2\, \lambda\, a_\ell}\, Q,\\
p_k&=&\Lambda(K,\hat{K})\,\ph_k,\;\;\;[k\,\neq\,(\ell,j)] 
\end{eqnarray*}
where $\Lambda$ is defined according to Eqn. (\ref{eq:LTini}), and where
\begin{\eqn*}
\hat{K}\,=\,Q-P_\ell,\;K\,=\,Q-p_\ell.
\end{\eqn*}

\subsubsection{The subprocess: $\boldsymbol{\gamma^* \,\to\, q(\hat p_1)\, \bar q(\hat p_2)\,g(\hat p_3)\,      g(\hat p_4)}$}
For this process, $ggg,\,qqg\,$ collinear as well as interference terms need to be taken into account; these are given by Eqns. (\ref{eq:dggg_fin}), (\ref{eq:dqqg_fin}), and (\ref{eq:Dif_fin}) respectively. We have for the collinear subtraction terms
\begin{eqnarray*}
\D^{\text{coll}}_{qqg;\ell j}&=&\frac{4\,\pi\,\al_s}{y\,(p_\ell\cdot Q)}\,C_F\,\left\{\frac{(\lambda-1+y)^{2}+4\,y}{4\,\lambda}\,F_\text{eik}\,+\,\frac{1}{2}\,z\,\left[ 1+y+\lambda\right]\right\},\\
\langle \nu|\D^{\text{coll}}_{ggg;\ell j}|\nu'\rangle\,\mathcal{T}^{\nu \nu'}&=&-2\,\pi\,\al_s\,C_A\,\frac{1}{(\ph_{\ell}
  \cdot \ph_{j})^{2}}\,\left[2-\Delta(2-\Delta)\right]\,\frac{1}{x_1^2+x_2^2}|\M_3(p_1,p_2,p_3)|^2\,\times\\
&&\Bigg\{\left[ \frac{2}{Q^2} \lb 2 (k_\perp\cdot p_1) (k_\perp \cdot p_2)-\frac{1-x_1}{1-x_2} (k_\perp\cdot p_1)^2 -\frac{1-x_2}{1-x_1} (k_\perp\cdot p_2)^2  \rb\right.\\
&&\left. \,+\,k_\perp^2\,\lb 1+\frac{x_1^2}{2}+\frac{x_2^2}{2}-x_1-x_2 \rb \right]  \Bigg\},
\end{eqnarray*}
with $k_\perp$ and $\Delta$ given by Eqns. (\ref{eq:kt}) and (\ref{eq:delta}). In total, we need to consider the following combinations:
\begin{\eqn*}
\D^{\text{coll}}_{qqg;1 3 },\,\D^{\text{coll}}_{qqg;1 4 },\,\D^{\text{coll}}_{qqg;2 3 },\,\D^{\text{coll}}_{qqg;2 4 },\,\D^{\text{coll}}_{ggg;3 4 },\,\D^{\text{coll}}_{ggg;4 3};
\end{\eqn*}
note especially that both combinations $(3,4),(4,3)$ need to be taken into account in the contributions from $ggg$ splittings.\\

The interference terms are given by
\beq 
{\cal D}^\text{if}_{\ell,j;k}\,=\Delta W_{\ell   j}(k)\,=\,4\,\pi\,\as\, C_{\ell k}\,
\frac{2\, \hat p_\ell\cdot\hat p_k\, \hat p_\ell\cdot\hat Q } 
{\hat p_\ell\cdot\hat p_j\,
\left(\hat p_j\cdot\hat p_k\, \hat p_\ell\cdot\hat Q+\hat p_\ell\cdot\hat p_j\,\hat p_k\cdot\hat Q \right)},
\eeq
where we have the following contributions:
\begin{\eqn*}
{\cal D}^\text{if}_{1,3;2},\,{\cal D}^\text{if}_{1,3;4},\,{\cal D}^\text{if}_{1,4;2},\,{\cal D}^\text{if}_{1,4;3},{\cal D}^\text{if}_{2,3;1},\,{\cal D}^\text{if}_{2,3;4},\,{\cal D}^\text{if}_{2,4;1},\,{\cal D}^\text{if}_{2,4;3},{\cal D}^\text{if}_{3,4;1},\,{\cal D}^\text{if}_{3,4;2},\,{\cal D}^\text{if}_{4,3;1},\,{\cal D}^\text{if}_{4,3;2}.
\end{\eqn*}
We want to emphasize again that there is {\sl only one mapping} required for each $(\ell,j)$ pairing, \ie we only have 5 independent mappings for the 12 subtraction terms listed above\footnote{The Catani Seymour prescription \cite{Catani:1996vz} requires 10 different mappings for this contribution.}. 
The subtracted cross section is then given by
\begin{alignat}{31}
\sigma^{R-A}\,=\,\int_4\,&\left[d\sigma^R \,-\, d\sigma^A\right] \notag \\
=\,\int \,dPS_4\,
&\left\{\,\frac{}{}
\left|{\cal M}_4(\hat p_1,\hat p_2,\hat p_3,\hat p_4)\right|^2\,F_4(\hat p_1,\hat p_2,\hat p_3,\hat p_4)   \right.\notag \\
&\left.-\langle\, 1,2,3\,|\,\D_{13}+\D_{14}+\D_{23}+\D_{24}+\D_{34}+\D_{43}\,|\,1,2,3 \, \rangle F_3(p_1,p_2,p_3) \,\frac{}{}\right\}\notag\\
\end{alignat}
with the spin-averaged matrix element
\begin{alignat}{31}
\frac{1}{4}\left|{\cal M}_4(\hat p_1,\hat p_2,\hat p_3,\hat p_4)\right|^2
&=
\sigma_0\,\left(\frac{\as}{2\,\pi}\right)^2\,C_F\,4\,\lb 4\,\pi \rb^5 \times\notag\\
&
\,
\left\{(A+B+C)+(1\leftrightarrow2)+(3\leftrightarrow4)+
(1\leftrightarrow2, 3\leftrightarrow4)\right\}
\end{alignat}
The quantities $A,B,C$ are given in Appendix B of \cite{Ellis:1980wv}.
\subsubsection{The subprocess: $\boldsymbol{\gamma^* \,\to\, q(\hat p_1)\, \bar q(\hat p_2)\,q(\hat p_3)\,\bar q(\hat p_4) }$}
This process does not contain any soft/ interference singularities, and all subtraction terms are given by Eqn. (\ref{eq:dgqq_fin}). We need to keep track of the mother parton's helicity in the subtraction term and obtain
\begin{eqnarray*}
\lefteqn{\langle \nu|{\cal D}_{gqq;\ell j}|\nu'\rangle\,\mathcal{T}^{\nu\,\nu'}\,=\,4\,\pi\,\al_s\,T_R\,\frac{1}{\ph_\ell\cdot\ph_j}\,|\M_3(p_1,p_2,p_3)|^2\,\times}\\
&&\Bigg\{1\,+\,\frac{2}{x_1^2+x_2^2}\,\frac{1}{\ph_\ell\cdot\ph_j}\,\left[ \frac{2}{Q^2}\,\lb 2\, (k_\perp\cdot\,p_1)\,(k_\perp \cdot p_2)-\frac{1-x_1}{1-x_2} (k_\perp\cdot\,p_1)^2 -\frac{1-x_2}{1-x_1} (k_\perp\cdot\,p_2)^2  \rb\right.\\
&&\left. \,+\,k_\perp^2\,\lb 1+\frac{x_1^2}{2}+\frac{x_2^2}{2}-x_1-x_2 \rb \right]  \Bigg\},
\end{eqnarray*}
with $k_\perp$ and $k_\perp^2$ given by Eqns. (\ref{eq:kt}) and (\ref{eq:ktsq}) respectively.

We have to consider the following combinations
\bea
{\cal D}_{gqq;12},\,{\cal D}_{gqq;14},\,{\cal D}_{gqq;32},\,{\cal D}_{gqq;34}
\eea
and get
\begin{alignat}{31}
\sigma^{R-A}\,=\,\int_4\,&\left[d\sigma^R \,-\, d\sigma^A\right] \notag \\
=\,\int \,dPS_4\,
&\left\{\,\frac{}{}
\left|{\cal M}_4(\hat p_1,\hat p_2,\hat p_3,\hat p_4)\right|^2\,F_4(\hat p_1,\hat p_2,\hat p_3,\hat p_4)   \right.\notag \\
&\left.-\langle\, 1,2,3\,|\,\D_{13}+\D_{14}+\D_{32}+\D_{34}|\,1,2,3 \, \rangle F_3(p_1,p_2,p_3) \,\frac{}{}\right\}\notag\\
\end{alignat}
with the spin-averaged matrix element given by
\begin{alignat}{31}
\frac{1}{4}\left|{\cal M}_4(\hat p_1,\hat p_2,\hat p_3,\hat p_4)\right|^2&=
\sigma_0\,\left(\frac{\as}{2\,\pi}\right)^2\,C_F\,4\,\lb 4\,\pi\rb^5
\, \times\notag\\
&
\left\{(D+E)+(1\leftrightarrow2)+(3\leftrightarrow4)+
(1\leftrightarrow2, 3\leftrightarrow4)\right\}
\end{alignat}
The quantities $D$ and $E$ are given in Appendix B of \cite{Ellis:1980wv}. 

\subsection{Integrated subtraction terms}

\subsubsection{Collinear integrals}

The collinear integrals involve the  $gq\bar{q}$, $qqg/{\bar q\bar qg}$ and $ggg$ splittings:
\beq
I_{\textrm{coll}}\,=\,I_{\textrm{coll}}(gq\bar{q})\,+\,
I_{\textrm{coll}}(qqg)\,+\,I_{\textrm{coll}}(ggg).
\eeq
After summing over all contributions, we obtain from Eqns. (\ref{eq:vgqq_fin}), (\ref{eq:vqqg_fin}) and (\ref{eq:vggg_fin})
\begin{eqnarray}
\lefteqn{ \langle 1, 2,3\left|\,I_{\textrm{coll}}(gq\bar{q}) \, \right|1, 2,3\rangle_B  \, =\,}\nn \\
&&       
\left|{\cal M}_3(p_1,p_2,p_3)\right|^2
\frac{\as}{2\,\pi}\,T_R\,\left(\frac{4\,\pi\,\mu^{2}}{Q^2}\right)^\vareps\,
\frac{1}{\Gamma(1-\vareps)}\,n_f\,\times\nn\\
&&\left[-\frac{2}{3\,\vareps}\,-\,\frac{16}{9}\,+\,\frac{2}{3}\,\left[(a_3-1)\,\ln(a_3-1)-(a_3+1)\,\ln\,a_3\right]\right],
                                               \nn \\
&&\nn \\
\lefteqn{\langle 1, 2,3\left|\,I_{\textrm{coll}}(qqg) \, \right|1, 2,3\rangle_A   \,  = \,}\nn\\
&&\left|{\cal M}_3(p_1,p_2,p_3)\right|^2
\frac{\as}{2\,\pi}\,C_F\,\left(\frac{4\,\pi\,\mu^{2}}{Q^2}\right)^\vareps\,
\frac{1}{\Gamma(1-\vareps)}\,\times\,\nn \\
&&\frac{1}{2}\sum_{\ell=1,2}\Bigg\{-\frac{1}{\vareps}+\,4\,I_{3}(a_\ell)\,-\,\ln\,a_\ell  +\frac{1}{2} \left[(9-7 a_\ell) (a_\ell-1) \ln (a_\ell-1)+a_\ell (7 a_\ell-16) \log (a_\ell)\right.\nn\\
&&\left. -7 \log (\ymax(a_\ell))-a_\ell (2 \ymax(a_\ell)+7)-7 \ymax(a_\ell)-4\right]\Bigg\}               \nn \\
&&\nn \\
\lefteqn{\langle 1, 2,3\left|\,I_{\textrm{coll}}(ggg) \, \right|1, 2,3\rangle_A  \,   \rangle\,=\,}\nn\\
&&           \left|{\cal M}_3(p_1,p_2,p_3)\right|^2
\frac{\as}{2\,\pi}\,C_A\,\left(\frac{4\,\pi\,\mu^{2}}{Q^2}\right)^\vareps\,
\frac{1}{\Gamma(1-\vareps)}\,\times\,\nn\\
&&\left\{-\frac{1}{6\,\vareps}\,-\,\frac{4}{9}\,+\,\frac{1}{6}\,\left[(a_3-1)\,\ln(a_3-1)\,-\,(a_3+1)\ln\, a_3\right]\,+\,I_\text{fin}(a_3)\right\},
\end{eqnarray}
where all symmetry factors have already been taken into account. Here, $I_\text{fin}$ and $I_3$ are given by Eqns. (\ref{eq:ifin}) and (\ref{eq:i3}) respectively and need to be evaluated numerically.

\subsubsection{Soft  integrals}

For a specific emitter/ spectator pair $(p_\ell,p_k)$, we obtain from Eqn. (\ref{eq:vif})
\begin{alignat}{31}
&I_{\textrm{soft},\ell k}\,=\,\frac{
4\,\pi\,\as}{2}\,C_{\ell k}\,\mu^{2\vareps}
\int d\xi_p\, \Delta W_{\ell k} =\notag\\
&
\frac{\as}{2\,\pi}\,C_{\ell k}\, 
\left(\frac{4\pi\mu^{2}}{Q^2}\right)^\vareps\frac{1}{\Gamma(1-\vareps)}\times
\Bigg\{\frac{1}{2\,\vareps^{2}}\,+\,\frac{1}{\vareps}\,\left[1\,+\,\frac{1}{2}\,\ln\left[ (\tilde{a}^{(\ell k)}_{0}+a_\ell)\,a_\ell\right]\right]\, -\,\frac{\pi^{2}}{6}\,+\,3\notag\\
&\,+\,\frac{1}{\pi}\left[\tilde{I}^{(b)}_\text{fin}\lb\frac{\tilde{a}^{(\ell k)}_{0}}{a_\ell}\rb\,+\,I^{(d)}_\text{fin}(\tilde{a}^{(\ell k)},a_\ell)\,+\,I^{(e)}_\text{fin}(a_\ell)\right] +\,\frac{1}{4}\,\ln^2\left[ (\tilde{a}^{(\ell k)}_0+a_\ell)\,a_\ell\right]\notag\\
&\,-\ln\,2\,\ln\lb \frac{\tilde{a}^{(\ell k)}_0+a_\ell}{a_\ell}\rb+2\,\ln\,a_\ell 
\Bigg\},
\end{alignat}
where 
\begin{\eqn*}
\tilde{I}_\text{fin}^{(b)}(b)\,\equiv\,I_\text{fin}^{(b)}(b)-\pi\left[\ln\,2\,\ln(1+b)\,+\,\frac{1}{4}\,\ln^2(1+b) \right].
\end{\eqn*}
The additional factor $\frac{1}{2}$ in the integrated subtraction terms stems from Eqn. (\ref{eq:v_ikp}) and accounts for the different symmetry factors and combinatorics of process (A).
Soft interference terms only appear for this process, where the following emitter/ spectator pairs need to be taken into account
\begin{\eqn}\label{eq:softcomb}
2\,\left[ (1,2)+(1,3)+(2,1)+(2,3)+(3,1)+(3,2) \right];
\end{\eqn}
the factor $2$ arises as \eg $(\ph_\ell,\ph_j,\ph_k)\,=\,\left[(\ph_1\,\ph_3,\,\ph_2),(\ph_1\,\ph_4,\,\ph_2)\right]$ are mapped to the same Born-type kinematics $(p_\ell,p_k)\,=\,(p_1,p_2)$, and similar relations hold for the other contributions.

\subsubsection{Finite parts}

Combining the one-loop matrix element \eq (\ref{loopthreejets31}) with the integrated subtraction terms, all poles in $\vareps$ cancel. The leftover finite parts are
\begin{alignat}{31}
\label{softintegralsFiniteparts131}
&
< 1, 2,3\left|\,I_\text{coll} \, \right|1, 2,3>_{\textrm{finite}}
\,=\,
\,
\left|{\cal M}_3(p_1,p_2,p_3)\right|^2
\,\frac{\as}{2\,\pi}\,\,\times
\notag\\
&\Bigg\{\left(4\,n_f\,T_R+C_A \right)\,\left[-\frac{4}{9}\,+\,\frac{1}{6}\,\left[(a_3-1)\,\ln(a_3-1)-(a_3+1)\,\ln\,a_3\right]  \right]+C_A\,I_\text{fin}(a_3)\notag\\
&+\,\frac{C_F}{2}\sum_{\ell=1,2}\Bigg[4\,I_{3}(a_\ell)\,-\,\ln\,a_\ell  +\frac{1}{2} \left[(9-7 a_\ell) (a_\ell-1) \ln (a_\ell-1)+a_\ell (7 a_\ell-16) \log (a_\ell)\right.\notag\\
&\hspace{25mm}\left. -7 \log (\ymax(a_\ell))-a_\ell (2 \ymax(a_\ell)+7)-7 \ymax(a_\ell)-4\right]\Bigg]\Bigg\}      
\end{alignat}
and 
\begin{alignat}{31}
\label{softintegralsFiniteparts132}
&
< 1, 2,3\left|\,I_\text{soft} \, \right|1, 2,3>_{\textrm{finite}}
\,=\,
\,
\left|{\cal M}_3(p_1,p_2,p_3)\right|^2
\,\frac{\as}{2\,\pi}\,\,\times
\notag\\
&\Bigg\{\left(2\,C_F+C_A \right)\,\left[-\frac{\pi^2}{3}+6  \right]+2\,\sum_{\ell=1,2,3}\,C_\ell^2\,\left[\frac{1}{\pi} I^{(e)}_\text{fin}(a_\ell)+2\,\ln\,a_\ell \right]\notag\\
&-2\,\lb C_A-2\,C_F \rb\,\left[ \frac{1}{\pi}\,\tilde{I}^{(b)}_\text{fin}\lb \frac{\tilde{a}_0^{(1,2)}}{a_1}\rb\,-\,\ln\,2\,\ln\lb\frac{\tilde{a}_0^{(1,2)}}{a_1} +1 \rb  \right]\notag\\
&+2\,C_A\,\sum_{\ell=1,2}\left[ \frac{1}{\pi}\,\tilde{I}^{(b)}_\text{fin}\lb \frac{\tilde{a}_0^{(\ell,3)}}{a_\ell}\rb\,-\,\ln\,2\,\ln\lb\frac{\tilde{a}_0^{(\ell,3)}}{a_\ell} +1 \rb  \right]\notag\\
&+\sum_{(\ell,k)}C_{\ell k}\,\left[\frac{2}{\pi} I^{(d)}_\text{fin}\,\lb \tilde{a}^{(\ell,k)},a_\ell \rb+\frac{1}{2}\,\ln^2\left[\lb \tilde{a}_0^{(\ell,k)}+a_\ell \rb a_\ell \right]\right]
\Bigg\},      
\end{alignat}
where the sum in the last line goes over all possible combinations as given in Eqn. (\ref{eq:softcomb}) (the factor 2 is already accounted for), and where we made use of several symmetries\footnote{One useful relation is \eg \begin{\eqn*} \frac{\tilde{a}_0^{(m, n)}}{a_n}\,=\,\frac{\tilde{a}_0^{(n, m)}}{a_m}.
\end{\eqn*}}. Further simplifications for the interference term finally render
\begin{alignat}{32}
&< 1, 2,3\left|\,I_\text{soft} \, \right|1, 2,3>_{\textrm{finite}}
\,=\,
\,
\left|{\cal M}_3(p_1,p_2,p_3)\right|^2
\,\frac{\as}{2\,\pi}\,\,\times
\notag\\
&\Bigg\{\left(2\,C_F+C_A \right)\,\left[-\frac{\pi^2}{3}+6  \right]+2\,\sum_{\ell=1,2,3}\,C_\ell^2\,\left[\frac{1}{\pi} I^{(e)}_\text{fin}(a_\ell)+2\,\ln\,a_\ell \right]\notag\\
&  +\frac{1}{2}C_A\,\left\{2\,\left[
  \ln\,a_1\,\ln\,a_2-\ln\,a_3\,\ln\lb \frac{a_1\,a_2}{a_3}
  \rb\right]+\,\ln^2\,y_{13}\,+\ln^2\,y_{23}\,-\ln^2\,y_{12}\right\} \notag\\
&+\,C_F\,\left[\ln^2\lb\frac{a_1}{a_2} \rb\,+\,\ln^2y_{12}   \right]\,+2\,\ln\,2\,\left[2\,C_F\,\ln\,\lb y_{12}\,a_1\,a_2\rb-C_A\,\ln\frac{y_{12}}{y_{13}y_{23}a_3^2}  \right]\notag\\
&+\frac{2}{\pi}\left[C_A \sum_{\ell=1,2} \tilde{I}^{(b)}_\text{fin}\lb \frac{1}{y_{\ell 3}a_\ell a_3}\rb - \lb C_A-2 C_F \rb \tilde{I}^{(b)}_\text{fin}\lb \frac{1}{y_{12} a_1 a_2}\rb  +\sum_{(\ell,k)}C_{\ell k}\, I^{(d)}_\text{fin} \lb \tilde{a}^{(\ell,k)},a_\ell \rb\right]
\Bigg\}      
\end{alignat}

\subsubsection{Final expressions}
The finite parts of one-loop matrix element are given by
\begin{alignat}{53}
\label{one-loopmatrixelementFiniteparts131}
&\left|{\cal M}_V(p_1,p_2,p_3) \right|^2_{\textrm{finite}} \,=\, 
\left|{\cal M}_3(p_1,p_2,p_3)\right|^2 \, \frac{\as}{2\pi}\, \times
\notag \\
&
\left\{\frac{}{} \frac{1}{2} \left[\frac{}{} \left(C_A-2\,C_F \right) \ln^2 y_{12} -
C_A \left(\ln^2 y_{13} + \ln^2y_{23} \right) \frac{}{}\right]
+  \frac{\pi^2}{2}\,(2\,C_F+C_A) - 8\, C_F \frac{}{}\right\}   \notag \\
&+ \frac{\as}{2\pi}\,\left[\frac{}{} F(y_{12},y_{13},y_{23}) +{\cal O}(\vareps)\frac{}{} \right] 
\end{alignat}

If we combine the one-loop matrix element with the integrated subtraction terms, poles in $\vareps$ exactly cancel, leading to finite results:
\begin{alignat}{53}
&\sigma^{V+A}= \int_3\, \left[d\sigma^V+\int_1d\sigma^A\right] \notag \\
&=\int dPS_3 \Bigg\{
\langle 1, 2,3\left|I_{\textrm{soft}}\right|1, 2,3\rangle +  \langle 1, 2,3\left|\,I_{\textrm{coll}} \right|1, 2,3\rangle
 + \left|{\cal M}_V(p_1,p_2,p_3) \right|^2  \Bigg\} F_{3} (p_1,p_2,p_3)\notag \\
&=\int dPS_3\Bigg\{
\langle 1, 2,3\left|I_{\textrm{soft}}  \right|1, 2,3\rangle_{\textrm{fin}}+
\langle 1, 2,3\left|I_{\textrm{coll}}  \right|1, 2,3\rangle_{\textrm{fin}}  
 + \left|{\cal M}_V(p_1,p_2,p_3) \right|^2_{\textrm{fin}} \Bigg\} F_3 (p_1,p_2,p_3)\notag\\
\end{alignat}
where
\begin{alignat}{54}
&\langle 1, 2,3\left|I_{\textrm{soft}}  \right|1, 2,3\rangle_{\textrm{fin}}+
\langle 1, 2,3\left|I_{\textrm{coll}}  \right|1, 2,3\rangle_{\textrm{fin}}  
 + \left|{\cal M}_V(p_1,p_2,p_3) \right|^2_{\textrm{fin}}\notag\\
&\,=\left|{\cal M}_3(p_1,p_2,p_3)\right|^2 \, \frac{\as}{2\pi}\, \times\notag\\
&\Bigg\{(2\,C_F+C_A)\frac{\pi^2}{6}+2\,C_F\,+\,\frac{50}{9}\,C_A \,-\,\frac{16}{9}\,n_f\,T_R\,+\frac{7}{2}\,C_F\,\ln\lb a_1\,a_2 \rb\notag\\
& +\lb \frac{23}{6}\,C_A\, -\frac{2}{3}\,n_F\,T_R \rb\ln\,a_3 +\left(4\,n_f\,T_R+C_A \right)\,\left[\frac{1}{6}\,\left[(a_3-1)\,\ln(a_3-1)-a_3\,\ln\,a_3\right]  \right]\notag\\
&+C_A\,I_\text{fin}(a_3)+\,\frac{C_F}{2}\sum_{\ell=1,2}\Bigg[4\,I_{3}(a_\ell)\,+\frac{1}{2} \left[(9-7 a_\ell) (a_\ell-1) \ln (a_\ell-1)+a_\ell (7 a_\ell-16) \log (a_\ell)\right.\notag\\
&\hspace{25mm}\left. -7 \log (\ymax(a_\ell))-a_\ell (2 \ymax(a_\ell)+7)-7 \ymax(a_\ell)\right]\Bigg]\notag\\
&  +\frac{1}{2}C_A\,\left\{2\,\left[
  \ln\,a_1\,\ln\,a_2-\ln\,a_3\,\ln\lb \frac{a_1\,a_2}{a_3}
  \rb\right]\right\} +\,C_F\,\ln^2\lb\frac{a_1}{a_2} \rb\,\notag\\
&+2\,\ln\,2\,\left[2\,C_F\,\ln\,\lb y_{12}\,a_1\,a_2\rb-C_A\,\ln\frac{y_{12}}{y_{13}y_{23}a_3^2}  \right]+\frac{2}{\pi}\,\sum_{\ell=1,2,3}\,C_\ell^2\,I^{(e)}_\text{fin}(a_\ell)\notag\\
&+\frac{2}{\pi}\left[ C_A \sum_{\ell=1,2} \tilde{I}^{(b)}_\text{fin}\lb \frac{1}{y_{\ell 3}a_\ell a_3}\rb - \lb C_A-2\,C_F \rb \tilde{I}^{(b)}_\text{fin}\lb \frac{1}{y_{12} a_1 a_2}\rb  +\sum_{(\ell,k)}C_{\ell k} I^{(d)}_\text{fin} \lb \tilde{a}^{(\ell,k)},a_\ell \rb\right]
\Bigg\}\notag\\
&+\frac{\al_s}{2\,\pi}\,F(y_{12},y_{13},y_{23}).
\end{alignat}

\subsection{Results}
\begin{figure}
\begin{minipage}{0.49\textwidth}
\begin{center}
\includegraphics[width=\textwidth]{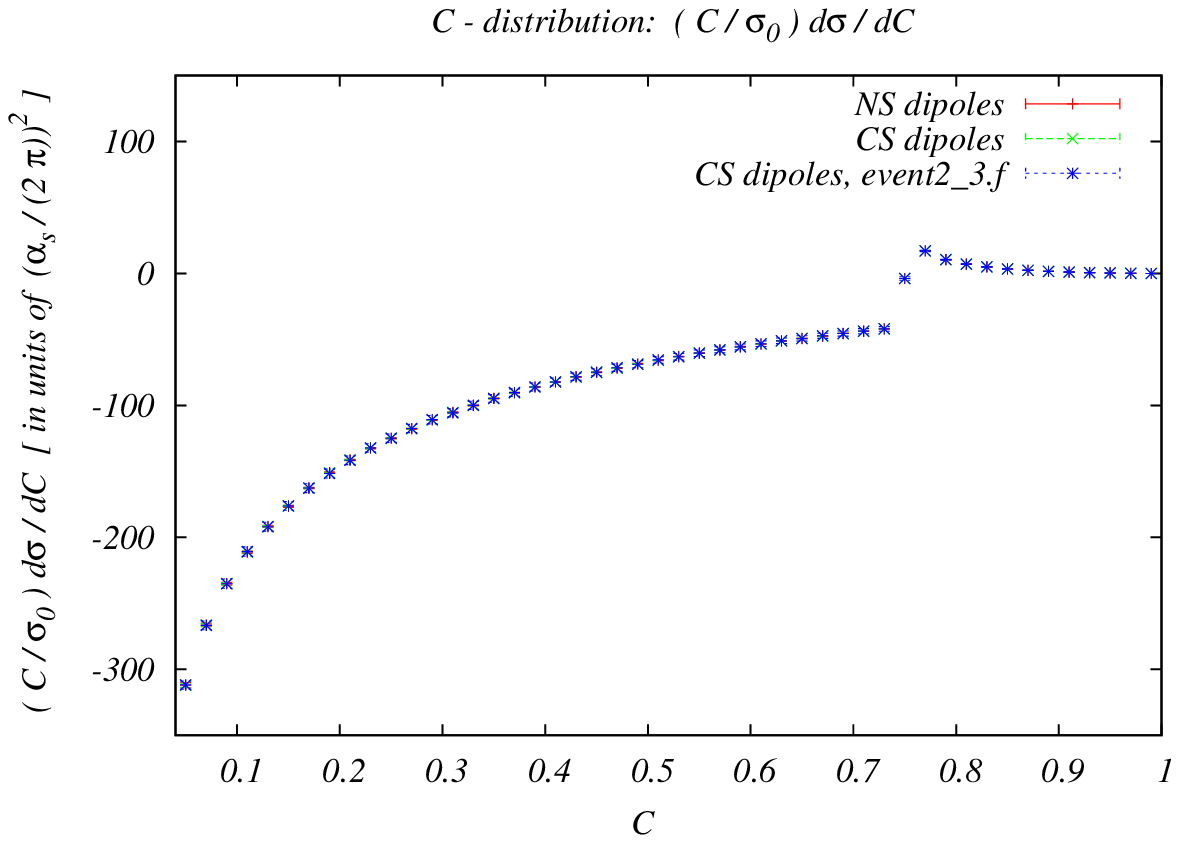}
\end{center}
\end{minipage}
\begin{minipage}{0.49\textwidth}
\begin{center}
\includegraphics[width=\textwidth]{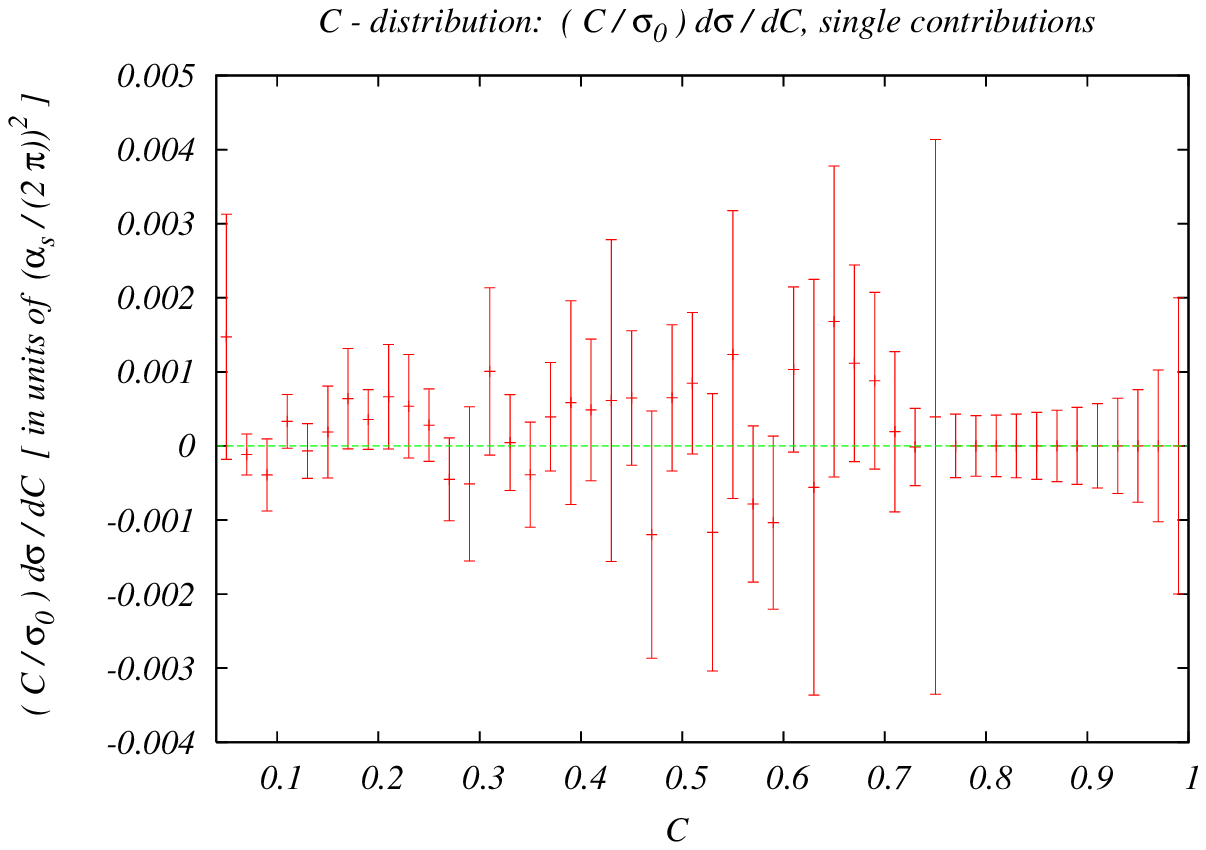}
\end{center}
\end{minipage}
\begin{minipage}{0.49\textwidth}
\begin{center}
\includegraphics[width=\textwidth]{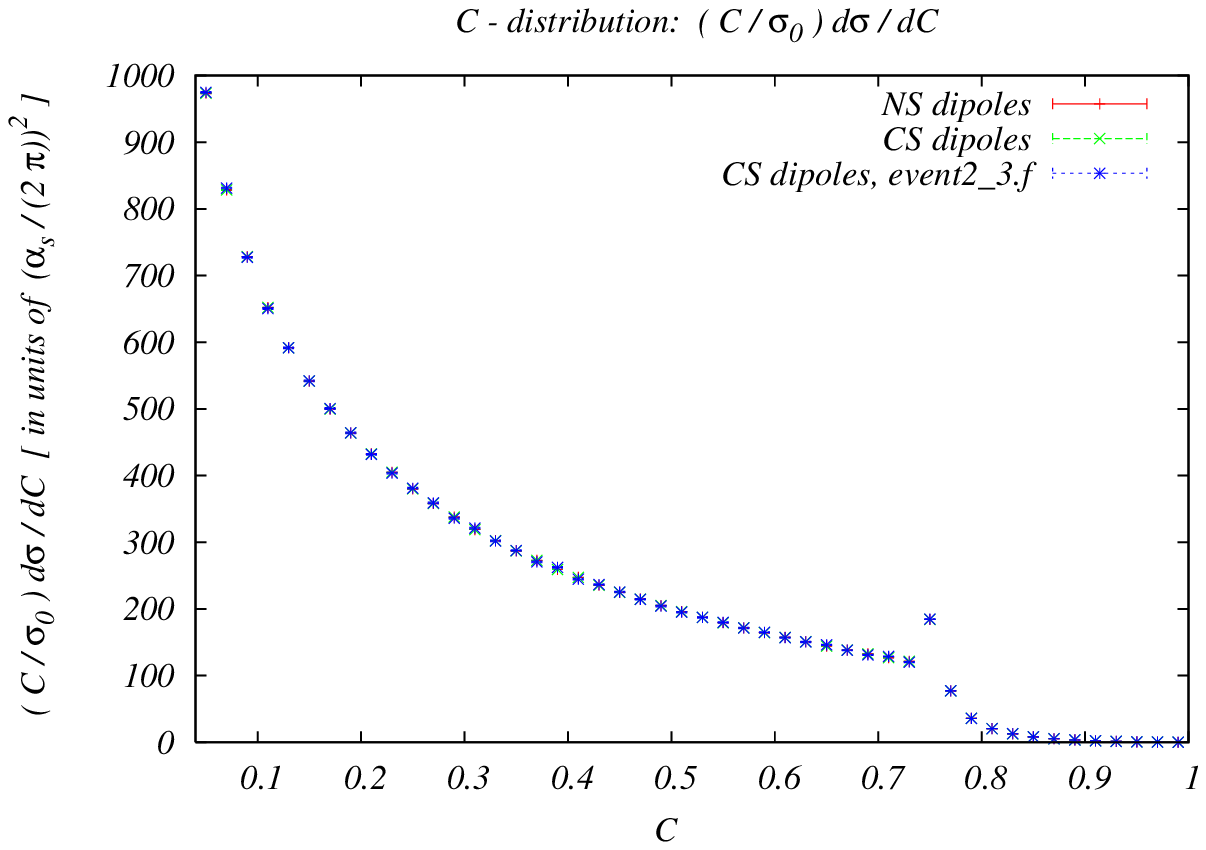}
\end{center}
\end{minipage}
\begin{minipage}{0.49\textwidth}
\begin{center}
\includegraphics[width=\textwidth]{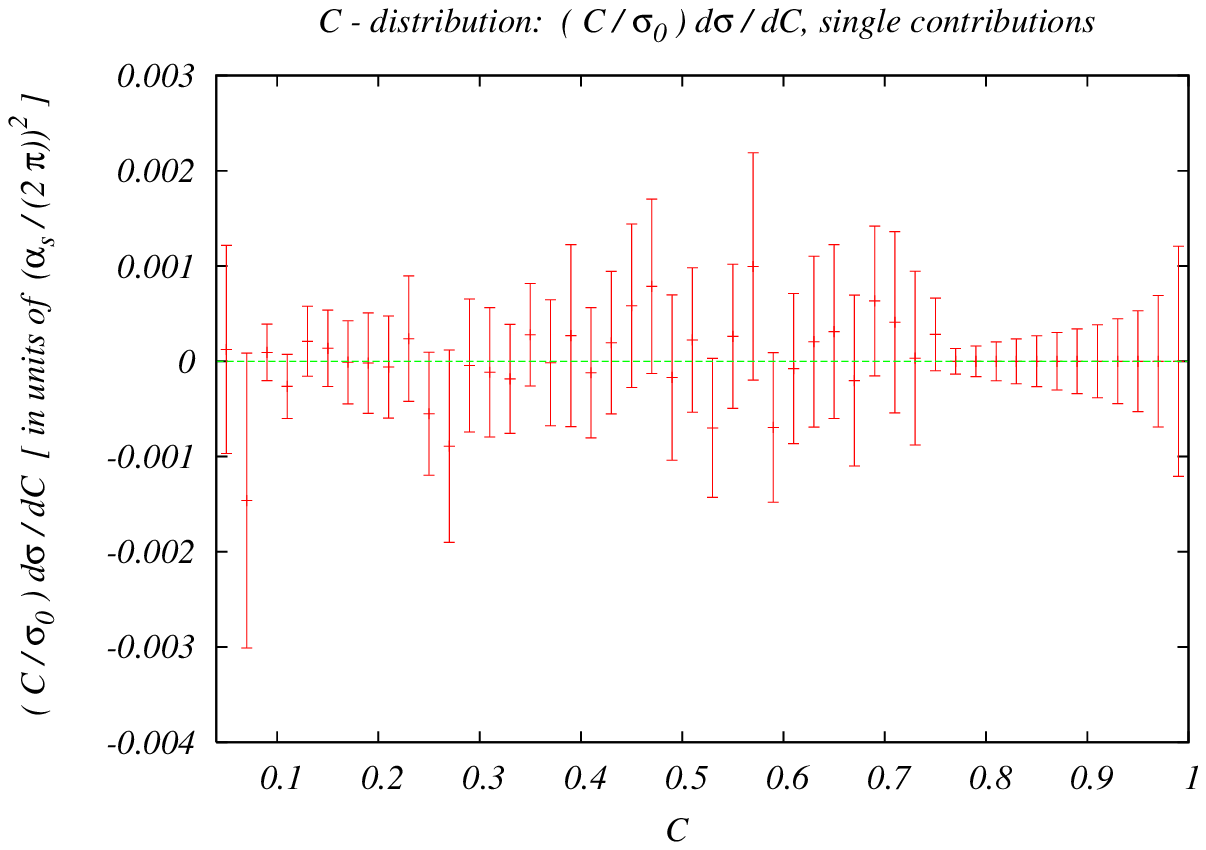}
\end{center}
\end{minipage}
\begin{minipage}{0.49\textwidth}
\begin{center}
\includegraphics[width=\textwidth]{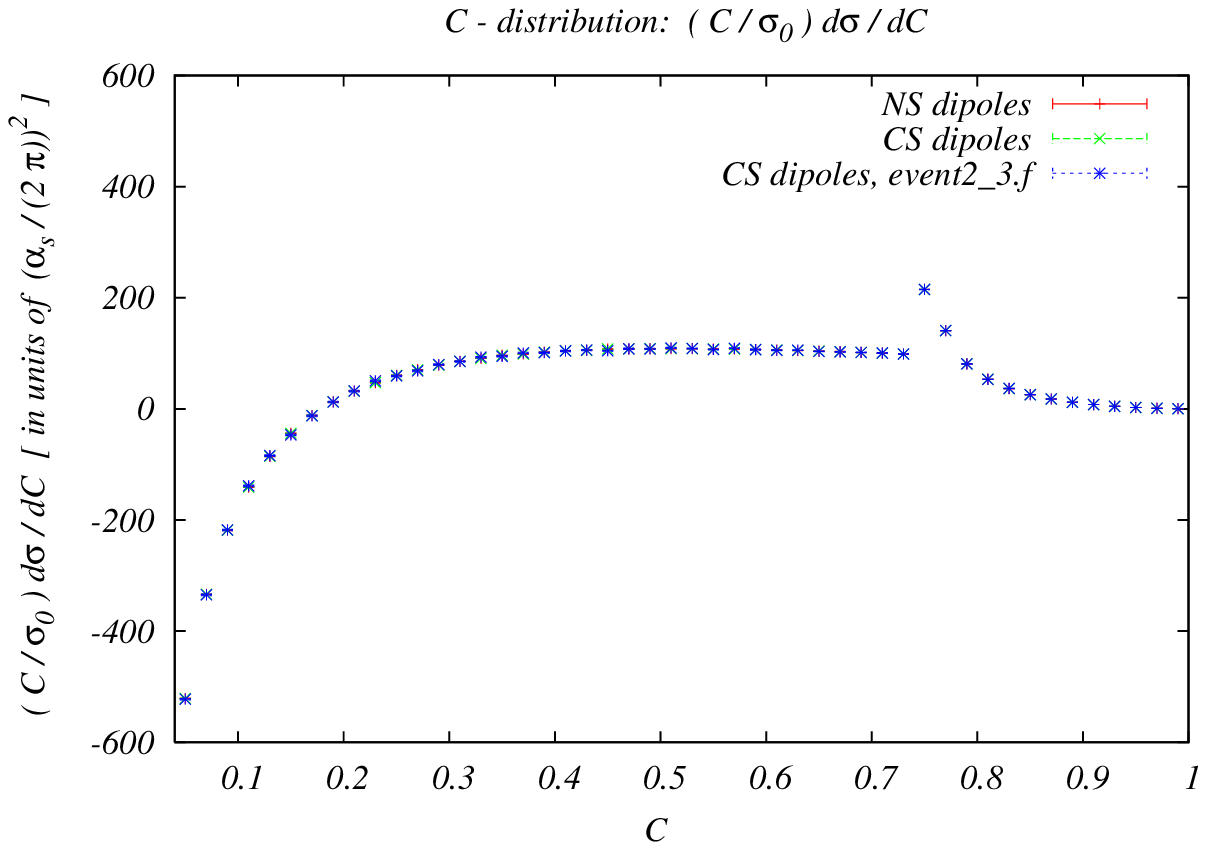}
\end{center}
\end{minipage}
\begin{minipage}{0.49\textwidth}
\begin{center}
\includegraphics[width=\textwidth]{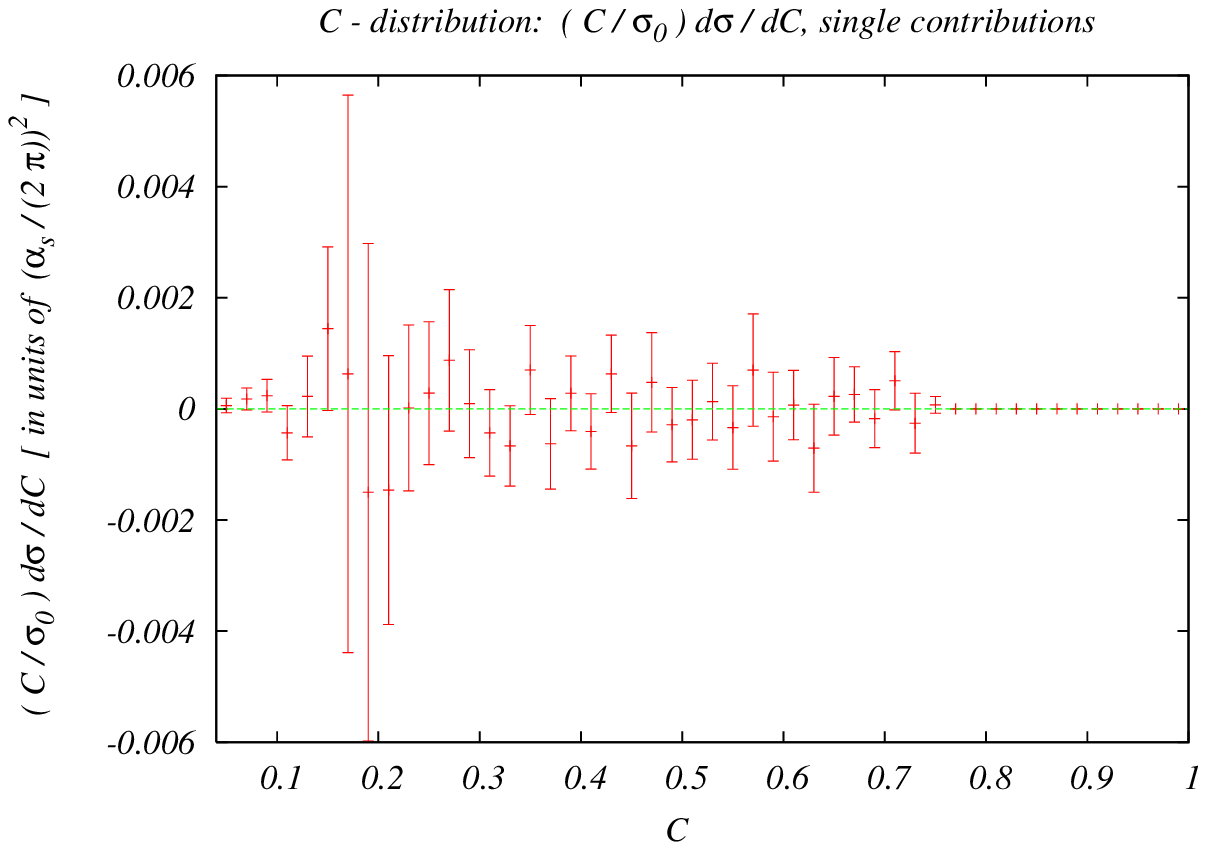}
\end{center}
\end{minipage}
\caption{\label{fig:a} {\sl From top to bottom:} Comparison of the contribution proportional to $N_C\,C_F\,n_f\,T_R,\;N_C^2\,C_F,\;N_C\,C_F^2$ for our scheme (NS) and Catani Seymour dipoles (CS). Shown is the differential distribution $\frac{C}{\sigma_0}\,\frac{d\sigma^\text{NLO}}{dC}$ in units of $\lb \frac{\al_s}{2\,\pi} \rb^2$, with C defined by Eqn.(\ref{eq:C}). {\sl Left:} Results for the implementation of our scheme and Catani Seymour subtraction terms from our private code as well as event2$\_$3.f \cite{event2f}. {\sl Right:} Relative difference between our implementation of the CS and NS subtraction terms. The results agree on the percent level and are consistent with zero within the integration errors. Large errors arise in regions where the absolute values of the differential distribution become small.} 
\end{figure}

\begin{figure}
\begin{center}
\includegraphics[width=0.75\textwidth]{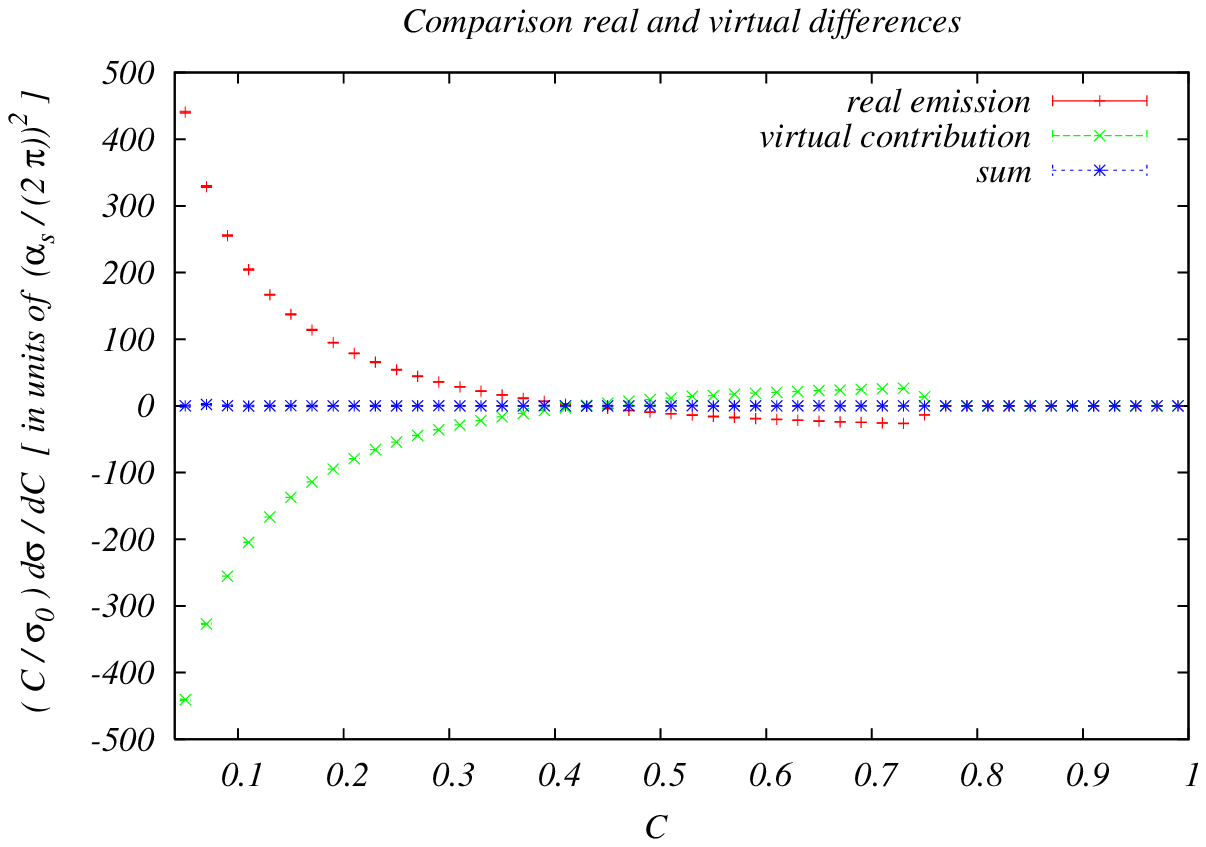}
\caption{\label{fig:difference} {\sl Differences} $\Delta_\text{CS-NS}$ for real emission (red, upper) and virtual (green, lower) virtual contributions, showing that especially for low $C$ values the contributions in the two schemes significantly differ. Adding up $\Delta^\text{real}+\Delta^\text{virt}$ gives 0 as expected.}
\end{center}
\end{figure}

We compared the implementation of our scheme with the results from \cite{Catani:1996jh,Catani:1996vz} as well as our own implementation of the Catani Seymour scheme. The subtraction terms for the latter are well known and will not be repeated here. To fulfill the jet-function requirements in Eqn. (\ref{eq:fjcond}), we chose the variable \cite{Ellis:1980wv},
\begin{\eqn}\label{eq:C}
C^{(n)}\,=\,3\,\left\{ 1-\sum_{i,j\,=\,1,\,i<j}^n\,\frac{s_{ij}^2}{(2\,p_i\cdot\,Q)\,(2\,p_j\cdot\,Q)}  \right\}
\end{\eqn}
which is infrared finite as required\footnote{A closer inspection of this variable shows that it contains singular regions which however are integrable \cite{Catani:1997xc}; we thank B. Webber for pointing this out.}. We numerically compared all different color contributions $N_C\,C_F^2,\,N_c\,C_F\,n_f\,T_R,\,N_C^2\,C_F$ separately, as well as the combined contributions. We set $n_f\,=\,5$ in our calculations. Figures \ref{fig:a} to \ref{fig:difference} show that we agree with results obtained using the Catani Seymour subtraction scheme on percent-level, and are consistent with 0, within the error bars, and thereby successfully validated our real emission subtraction terms as well as all integrated counterterms proposed in this paper.
\begin{figure}
\begin{center}
\includegraphics[width=0.75\textwidth]{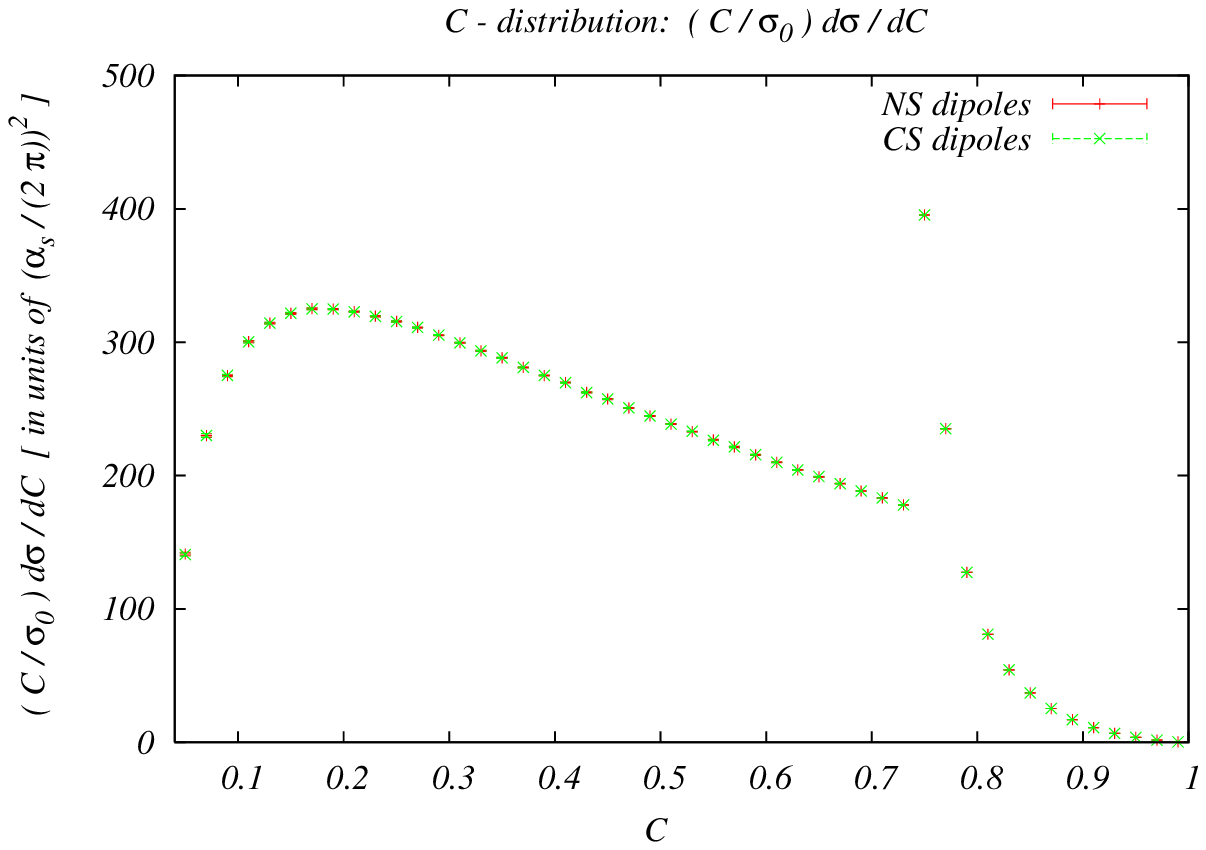}
\caption{Total result for differential distribution $\frac{C}{\sigma_0}\,\frac{d\sigma^\text{NLO}}{dC}$ using both NS (red) and CS (green) dipoles. The standard literature result obtained using the CS scheme is completely reproduced with the NS dipoles.}
\end{center}
\end{figure}
Results for integration as well as differential distributions have been obtained using routines from the Cuba library \cite{Hahn:2004fe}.

\clearpage

\section{Conclusion and Outlook}
In this work, we have extended the alternative subtraction scheme for NLO QCD calculations proposed in \cite{Robens:2010zr,Chung:2010fx,Robens:2011bz} to the case of an arbitrary number of massless partons in the final state. The scheme employs a momentum mapping that reduces the number of reevaluations of the underlying Born matrix element with respect to standard schemes \cite{Catani:1996vz,Catani:2002hc} \footnote{We want to note that the Frixione-Kunszt-Signer (FKS) subtraction scheme \cite{Frixione:1995ms} exhibits a scaling behaviour similar to our scheme. However, the two prescriptions differ in the treatment of phase space setup; in addition, no parton shower proposal exists using FKS splitting functions. We thank R. Frederix for helpful discussions regarding this point.}. Furthermore, the use of subtraction terms based on the shower splitting functions promises to facilitate the matching of parton-level NLO corrections with the improved parton shower\footnote{For an explicit discussion on this, see eg \cite{Hoeche:2011fd, Hoeche:2012fm}, where the authors additionally emphasize that in case of processes with subleading colour divergences, this choice naturally allows to sustain NLO accuracy of total and differential distributions after matching with the shower. We thank S. H\"oche for valuable comments regarding this.}. We provide formulae for the corresponding final state subtraction terms and their integrated counterparts. We validate our expressions by reproducing the literature results for the differential distribution of the $C$ parameter at NLO for three jet production at lepton colliders, where we find numerical agreement between results from the implementation of our scheme and two independent implementations of the Catani Seymour scheme on the (sub)percent level. Combining the results in this work with the discussion in \cite{Robens:2010zr,Chung:2010fx,Robens:2011bz} provides all formulae needed for a generic application of our scheme for massless emitters, and therefore concludes the discussion of the subtraction scheme in the massless case.\\

As argued in the Introduction, subtraction schemes can generically differ in the nonsingular structure of the dipole subtraction terms as well as the mapping between real emission and Born-type kinematics, which guarantees on-shellness and energy momentum conservation in both phase spaces. The scheme adopted here uses the whole remaining event as a spectator for the mapping, thereby leading to a scaling behaviour of Born reevaluations $\sim\,N^2/2$, where $N$ is the number of final state partons. However, this simplified mapping equally induces integrated subtraction terms with finite parts that exhibit a non-trivial dependence on the integration parameters of the unresolved one-parton phase space. In this work, we chose to evaluate these finite terms numerically, which leads to an increase of integration variables by two in the numerical implementation of the scheme. However, recently it was proposed \cite{Maestre:2012vp} to approximate similar finite terms by polynomial functions in the context of a next-to-next-to-leading order subtraction scheme\cite{Somogyi:2005xz,Somogyi:2006cz,Somogyi:2006da,Nagy:2007mn}. We therefore plan to make the finite remainders that appear in the integrated subtraction terms available either in form of approximating functions or a librarized grid interpolating between different input parameters for $a_\ell$. Further plans for future work include the extension to the massive scheme as well as the matching with the improved parton shower{\footnote{The scheme presented
  here has recently been implemented into the Helac NLO event Generator framework \cite{Bevilacqua:2011xh,Bevilacqua:2013taa}. }.

\newpage

\section*{Acknowledgements}
This research was partially supported by the DFG SFB/TR9
``Computational Particle Physics'', the DFG Graduiertenkolleg
``Elementary Particle Physics at the TeV Scale'', the Helmholtz
Alliance ``Physics at the Terascale'' and the BMBF. We would like to thank Zolt\'{a}n
Nagy, Dave Soper, Zolt\'{a}n Tr\'{o}cs\'{a}nyi, Michael Kr\"amer and Gabor Somogyi for many valuable
discussions, as well as Tobias Huber for help with integrations
including HypExp and Mike Seymour for help with his code for $ee\,\rightarrow\,3$ jets. In addition, C.H.C wants to thank Alexander M\"uck and Michael Kubocz for useful discussions. T.R. thanks Thomas Hahn for useful comments regarding the Cuba library, Stefan H\"oche and Rikkert Frederix for useful comments on NLO and parton shower matching and the FKS implementation in Madgraph respectively, as well as Christoph Gnendiger for helpful comments regarding the manuscript. T.R. equally acknowledges financial support by the SFB 676 "Particles, strings, and the early universe" during the final stages of this work.
\newpage

\begin{appendix}
\section{Splitting amplitudes}\label{app:splittings}
In Table \ref{tab:splitting}, we list the splitting amplitudes for final state
$q\bar{q}g$ splittings as given in \cite{Nagy:2007ty}.
\begin{table}
\begin{tabular}{cccccc}
\hline
\\
$\ell$&$f_\ell$&$\hat f_\ell$&$\hat f_{j}$\!\!\!\!&
$\displaystyle{v_\ell \times \frac{1}{\sqrt{4\pi\as}}}$
&colour \\
\\
$F$&
$q$&
$q$&
g&
$\displaystyle{
\varepsilon_\mu(\hat p_{j},\hat s_{j};\hat Q)^*\,
\frac{
\overline U({\hat p_\ell,\hat s_\ell})\gamma^\mu 
  [\s{\hat p}_\ell + \s{\hat p}_{j}] \s{n_\ell} U({p_\ell,s_\ell})}
{2p_\ell\!\cdot\! n_\ell\ [(\hat p_\ell + \hat p_{j})^2]}
}
$&
$t^a$\\
&&&&&\\
$F$&
g&
$q$&
$\bar q$&
$\displaystyle{
-
\varepsilon^\mu(p_{\ell},s_{\ell};\hat Q)
D_{\mu\nu}(\hat p_\ell + \hat p_{j},n_\ell)
\frac{
\overline U({\hat p_\ell,\hat s_\ell})\gamma^\nu V({\hat p_{j},\hat s_{j})}
}{(\hat p_\ell + \hat p_{j})^2}
}$&
$t^a$\\
&&&&&\\
\\  [10 pt]
\hline
\end{tabular}
\caption{Splitting amplitudes $v_\ell(\{\hat p, \hat f\}_{j},\hat
  s_{j},\hat s_{\ell},s_\ell)$ 
  involving a $q\bar{q}g$ splitting. We have removed a common factor 
  $\sqrt{4\pi\as}$. The label $\ell$ denotes final state indices $F = \{1,\dots,m\}$. 
  The lightlike vector $n_\ell$ is defined in Eqn.~(\ref{eq:nldef}). Taken from \cite{Nagy:2007ty}.}
\label{tab:splitting}
\end{table}
For triple gluon splittings, we have for the final state
\begin{equation}
\begin{split}
\label{eq:VggF}
v_\ell(\{\hat p, \hat f\}_{m+1},&\hat s_{j},\hat s_{\ell},s_\ell)
\\ & =
\frac{\sqrt{4\pi\as}}{2 \hat p_{j}\!\cdot\! \hat p_\ell}\, 
\varepsilon_{\alpha}(\hat p_{j}, \hat s_{j};\hat Q)^*
\varepsilon_{\beta}(\hat p_{\ell}, \hat s_l;\hat Q)^*
\varepsilon^{\nu}(p_{\ell}, s_\ell;\hat Q)
\\&\quad\times
v^{\alpha \beta \gamma}(\hat p_{j},\hat p_\ell,-\hat p_{j}-\hat p_\ell)\,
D_{\gamma\nu}(\hat p_\ell + \hat p_{j};n_\ell)\,.
\end{split}
\end{equation}
We use standard notation where
$U(p,s),\overline{U}(p,s),V(p,s),\overline{V}(p,s)$ denote spinors of
the fermions with a four-momentum $p$ and spin $s$, and
$\varepsilon_{\alpha}(p, s;Q)$ are the gluon polarisation vectors. The
$ggg$ vertex has the form \beq
\label{eq:vgg}
v^{\alpha \beta \gamma}(p_a, p_b, p_c)
= g^{\alpha\beta} (p_a - p_b)^\gamma
+ g^{\beta\gamma} (p_b - p_c)^\alpha
+ g^{\gamma\alpha} (p_c - p_a)^\beta\,.
\eeq
The transverse projection tensor $D_{\gamma\nu}(\hat p_\ell - \hat p_{j};n_\ell)$ is defined according to \eq (\ref{transverseprojectiontensor}).
The lightlike vector $n_\ell$ is given by
\begin{equation}
\label{eq:nldef}
n_\ell = 
Q
-\frac{Q^2}
{Q\!\cdot\! p_\ell
+ \sqrt{(Q\!\cdot\! p_\ell)^2 }}\
p_\ell
\;\;,  \ell \in \{1,\dots,m\}\;\;.
\end{equation}

\section{Incoming hadrons}\label{app:cts}
In case of processes with two initial-state hadrons $A$ and $B$ carrying momenta
$p_A$ and $p_B$, respectively, the calculation of the QCD cross sections must be convoluted with parton distribution functions $f_{i/I}(\eta_i,\mu^2_F)$ which depends on the factorization scale $\mu_F$:
\begin{alignat}{53}
\label{eq:sig_had}
\sigma(p_A,p_B) &=\,\sum_{a,\,b}\int_0^1d\eta_a\, f_{a/A}(\eta_a,{\mu_F^2})\int_0^1d\eta_b\,f_{b/B}(\eta_b,{\mu_F^2}) 
          \left[\sigma^{\text{LO}}_{ab}(p_a,p_b)+\sigma^{\text{NLO}}_{ab}(p_a,p_b,\mu_F^2)\right] 
\end{alignat}
where $p_a\,=\,\eta_a\,p_A$ and $p_b\,=\,\eta_b\,p_B$ are parton momenta, while $\eta_a$ and $\eta_b$ are the momentum fractions of the partons. In this case, additional collinear counterterms need to be added to the integrated subtraction terms\footnote{{\sl C.f.} \eg \cite{Ellis:1991qj}.}, and the parton level NLO contribution becomes
\begin{alignat}{53}
\label{eq:siglo_nlo}
\sigma^{\text{NLO}}_{ab}(p_a,p_b,\mu_F^2)&=\int_{m+1}d\sigma^R_{ab}(p_a,p_b)+
\int_{m}d\sigma^V_{ab}(p_a,p_b)+\int_{m}d\sigma^C_{ab}(p_a,p_b,\mu_F^2).
\end{alignat}
We then have
\begin{alignat}{53}
\label{eq:sig_nlo}
\sigma_{ab}^{\text{NLO}}(p_a,p_b,\mu_F^2)&=  \int_{m+1} \left[ d\sigma_{ab}^{R}(p_a,p_b) -
d\sigma_{ab}^{A}(p_a,p_b) \right]                    \notag \\
&+ 
\int_{m}\, \left[\int d\sigma^{V}_{ab}(p_a,p_b)+ \int_1d\sigma^{A}_{ab}(p_a,p_b)+d\sigma_{ab}^{C}(p_a,p_b,\mu_F^2)\right]_{\vareps=0},
\end{alignat}
with
\begin{alignat}{53}
\label{IKP2842010}
&\int_{m}  \left[ \int_{1} d\sigma^A_{ab}(p_a,p_b)+ d\sigma^C_{ab}(p_a,p_b,\mu_F^2) \right]    \notag \\
=&
\int_m  d\sigma_{ab}^{B}(p_a,p_b) \otimes { I}(\vareps)
+ \int_0^1 dx \int_m  d\sigma_{ab}^{B}(xp_a,p_b)\otimes
\left[ { K}^a(xp_a) +   { P}(x,\mu_F^2)  \right]                             \notag \\
+&  \int_0^1 dx \int_m d\sigma_{ab}^{B}(p_a,xp_b)\otimes
\left[ { K}^b(xp_b) +   { P}(x,\mu_F^2)  \right].  
\end{alignat}
This equation defines the insertion operators $I(\vareps),\,K(x),\,
P(x;\mu_{F}^2)$ on an integrated cross section level.
\eq(\ref{IKP2842010}) 
can be divided into two parts: the first part is the universal insertion
operator ${ I}(\vareps)$, which contains the complete singularity
structure of the virtual contribution and has LO kinematics. The second part consists of  the finite pieces that are left over after absorbing the
initial-state collinear singularities into a redefinition of the parton distribution functions at NLO. 
It involves an additional one-dimensional integration over the momentum fraction $x$ of an incoming parton 
with the LO cross sections and the $x$-dependent structure functions. \\

In the $\msbar$
scheme, the collinear counterterms are given by
\begin{alignat}{53}
\label{universalcollinearcounterterm_general}
\int_md\sigma^C_{ab}(p_a,p_b,\mu_F^2)&= \frac{\as}{2 \pi}\,
\frac{1}{\Gamma(1-\vareps)}\,
\sum_{c}\int_0^1 dx\int_m d\sigma_{cb}^{B}(xp_a, p_b)\,\frac{1}{\vareps}\,
\left( \frac{4 \pi \mu^2}{\mu_F^2} \right)^{\vareps}\, P^{ac}(x)   \notag \\
&+ \frac{\as}{2 \pi}\,
\frac{1}{\Gamma(1-\vareps)}\,
\sum_{c}\int_0^1 dx\int_m d\sigma_{ac}^{B}(p_a, xp_b)\,\frac{1}{\vareps}\,
\left( \frac{4 \pi \mu^2}{\mu_F^2} \right)^{\vareps}\, P^{bc}(x) .  
\end{alignat}
Here the $ P^{ab}(x)$ are the Altarelli-Parisi kernels in four
dimensions \cite{Altarelli:1977zs}, which are evolution
kernels of the DGLAP equation \cite{Gribov:1972ri,
  Lipatov:1974qm,Altarelli:1977zs,Dokshitzer:1977sg}, and describe the behaviour of parton splittings by giving the probability
of finding a parton of type $b$ with momentum fraction $x$ in a parton
of type $a$ in the collinear limit:
\begin{\eqn}
a(p)\,\longrightarrow\,b\lb x\,p+k_{\perp}+\,\mO(k_{\perp}^{2})\rb\,+\,c\,\lb (1-z)\,p-k_{\perp}+\,\mO(k_{\perp}^{2}) \rb.
\end{\eqn}
At leading order, the splitting functions are given by
\begin{eqnarray}
P^{qq}(x)&=&C_{F}\,\left[ \frac{1+x^{2}}{(1-x)_{+}}+\frac{3}{2}\,\delta(1-x)\right],\nn\\
P^{gq}(x)&=&T_{R}\,\left[x^{2}+(1-x)^{2} \right],\;T_{R}\,=\,\frac{1}{2},\nn\\
P^{qg}(x)&=&C_{F}\,\left[\frac{1+(1-x)^{2}}{x}\right],\nn\\
P^{gg}(x)&=&2\,C_{A}\,\left[ \frac{x}{(1-x)_{+}}\,+\,\frac{1-x}{x}+x\,(1-x) \right]+\delta(1-x)\,\frac{11\,C_{A}-4\,n_{f}\,T_{R}}{6}\,,\nn\\&&
\end{eqnarray}
where $n_{f}$ is the number of quark flavours in the theory. The $+$
distribution is defined in the standard way
\begin{\eqn}\label{eq:plusdef}
\int^{1}_{0}\,f(x)\,g_{+}(x)\,dx\,=\,\int^{1}_{0}\,g(x)\,(f(x)-f(1))\,dx\,=\,\int^{1}_{0}\,g(x)\,f(x)\,dx\,-\,f(1)\,\int^{1}_{0}g(x)\,dx
\end{\eqn}
for the convolution with a test function $f(x)$.
\section{Four-particle phase space}
In this section, we derive the parametrization that was used for the real emission phase space in Section \ref{Threejetproduction2011}. We use the standard notation for an $n$-parton phase space in four dimensions:
\begin{\eqn*}
d\Gamma_n\,=\,\prod_i\,\left[\frac{d^4p_i}{(2\pi)^4}\,\delta\lb p_i^2-m_i^2 \rb\right]\,\delta^{(4)}\lb p_\text{in}-\sum_i\,p_i \rb.
\end{\eqn*}
We build our parametrization from a successive chain with
\begin{\eqn*}
p_\text{in}\,\rightarrow\,p_{12}+p_{34},\,p_{ij}\,\rightarrow\, p_i+p_j,
\end{\eqn*}
where in the first step the on-shell condition for $p_{ij}$ needs to be replaced by a distribution of $s_{ij}\,=\,(p_i+p_j)^2$.
\subsubsection*{Generic massive two parton phase space, center-of-mass system}
For a generic massive two parton phase space, we use the following parametrization in the center-of-mass system
\begin{\eqn*}
d\Gamma_2\,=\,\frac{1}{32\,\pi^2}\,\frac{\sqrt{\lambda(s,m_1^2,m_2^2)}}{s}d\Omega_1\,\Theta\lb  \sqrt{s}-(m_1+m_2)\rb,
\end{\eqn*}
where
\begin{\eqn*}
p_{1}^0\,=\,\frac{s+m_1^2-m_2^2}{2\,\sqrt{s}},\,|\ora{p}_{1}|\,=\,\frac{\sqrt{\lambda(s,m_1^2,m_2^2)}}{2\,\sqrt{s}},\,p_2\,=\,p_\text{in}-p_1.
\end{\eqn*}
The $\Theta$ function arises from the conditions $\Theta(|\ora{p}_1|)\,\Theta(|\ora{p}_2|)$. $\lambda$ is defined as
\begin{\eqn*}
\lambda(x_1,x_2,x_3)\,=\,\sum_i\,x_i^2-2\,\sum_{i\,>\,j}x_{i}x_{j}.
\end{\eqn*}
\subsubsection*{Generic massless two parton phase space, non-center-of-mass system}
If we consider two partons in a non-center-of-mass system, with
\begin{\eqn*}
p_\text{in}\,=\,\lb \begin{array}{c} E_\text{in}\\0\\0\\ |\ora{p}_\text{in}| \end{array} \rb
\end{\eqn*}
\ie where the three vector of $p_\text{in}$ determines the z axis, we obtain for a two parton massless phase space
\begin{\eqn*}
d\Gamma_2\,=\,\frac{1}{16\,\pi^2\,|\ora{p}_\text{in}|} dp_1^0\,d\phi_1,
\end{\eqn*} 
where we have
\begin{\eqn}\label{eq:costh}
\cos\theta_1\,=\,\frac{1}{|\ora{p}_\text{in}|}\lb E_\text{in}-\frac{m^2_\text{in}}{2\,p_1^0} \rb
\end{\eqn}
and
\begin{\eqn*}
\frac{m^2_\text{in}}{2\,(E_\text{in}+|\ora{p}_\text{in}|)}\,\leq\,p^0_1\,\leq\,\frac{m^2_\text{in}}{2\,(E_\text{in}-|\ora{p}_\text{in}|)}
\end{\eqn*}
from the requirement that $|\cos\theta_1|\leq\,1$. In this derivation,
\begin{\eqn*}
p_{1,z}\,=\,p_1^0\,\cos\theta_1.
\end{\eqn*}
If the z-component of $p_\text{in}$ goes into the negative $z$-direction, $\cos\theta\,\rightarrow\,-\cos\theta$ in Eqn. (\ref{eq:costh}), and all other above relations still hold.
\subsubsection*{Generic four parton phase space with massless final states}
We use the generic expression
\begin{\eqn*}
d\Gamma_n\,\lb X\rightarrow\,\sum p_n \rb\,=\,d\Gamma_{X\rightarrow\,Y+Z}\,\frac{d m^2_X}{2\pi}\,\frac{d m^2_Y}{2\pi}\,d\Gamma_{X\rightarrow\,\sum p_o}\,d\Gamma_{Y\rightarrow\,\sum p_o'},
\end{\eqn*}
where $\sum p_o\,+\,\sum p_o'\,=\,\sum p_n$ is the sum over all $n$ outgoing particles. Using the expressions above as well as 
\begin{\eqn*}
y_{ij}\,=\,\frac{s_{ij}}{s},\,x_i\,=\,2\frac{p_i\cdot Q}{s},
\end{\eqn*}
we obtain for the four-parton phase space in the center-of-mass system of $p_\text{in}\,\equiv\,Q$:
\begin{\eqn*}
d\Gamma_4\,=\,\frac{s^2}{(4\,\pi)^6\,\sqrt{\lambda(1,y_{12},y_{34})}}\,dy_{12}\,dy_{34}\,dx_1\,dx_3\,d\phi_3,
\end{\eqn*}
with the four-vectors
\begin{eqnarray*}
&&p_{12}\,=\,\lb \begin{array}{c} E_X\\0\\0\\p_X\end{array} \rb,\;p_{34}\,=\,\lb \begin{array}{c} E_Y \\0\\0\\-p_X\end{array} \rb,\\
&&p_1\,=\, x_1\,\frac{\sqrt{s}}{2}\,\lb\begin{array}{c} 1\\ \sin\theta_1\\0\\ \cos\theta_1 \end{array} \rb,\;p_3\,=\, x_3\,\frac{\sqrt{s}}{2}\,\lb\begin{array}{c} 1\\ \sin\theta_3\,\cos\phi_3\\\sin\theta_3 \sin\phi_3\\ \cos\theta_3 \end{array} \rb,\\
&&p_2\,=\,p_{12}-p_1,\;p_4\,=\,p_{34}-p_3,
\end{eqnarray*}
and where 
\begin{eqnarray*}
&& E_X\,=\,\frac{\sqrt{s}}{2}\lb 1 + y_{12}- y_{34}\rb,\,p_X\,=\,\frac{\sqrt{s\,\lambda(1,y_{12},y_{34})}}{2},\\
&&E_Y\,=\,\sqrt{s}-E_X\,=\,\frac{\sqrt{s}}{2}\lb 1 + y_{34}- y_{12}\rb,\\
&&\cos\theta_1\,=\,\frac{1}{p_X}\lb E_X-\frac{m^2_X}{x_1\,\sqrt{s}} \rb\,=\,\frac{1}{\sqrt{\lambda(1,y_{12},y_{34})}}\,\lb 1+y_{12}-y_{34}-\frac{2\,y_{12}}{x_1} \rb,\\
&&\cos\theta_3\,=\,-\frac{1}{p_X}\lb E_Y-\frac{m^2_Y}{x_3\,\sqrt{s}} \rb\,=\,-\frac{1}{\sqrt{\lambda(1,y_{12},y_{34})}}\,\lb 1+y_{34}-y_{12}-\frac{2\,y_{34}}{x_3} \rb.\
\end{eqnarray*}
The integration boundaries are given by
\begin{eqnarray*}
&&y_{34}\,\leq\,\lb 1-\sqrt{y_{12}} \rb^2,\\
&&x_1^\text{min/max}\,=\,\frac{m_X^2}{\sqrt{s}\,\lb E_X\,\pm\,p_X \rb}\,=\,\frac{2\,y_{12}}{1-y_{12}+y_{34}\,\pm\,\sqrt{\lambda(1,y_{12},y_{34})}},\\
&&x_3^\text{min/max}\,=\,\frac{m_Y^2}{\sqrt{s}\,\lb E_Y\,\pm\,p_X \rb}\,=\,\frac{2\,y_{34}}{1-y_{34}+y_{12}\,\pm\,\sqrt{\lambda(1,y_{12},y_{34})}}.
\end{eqnarray*}
\section{Note on further possible scaling improvement}
As argued in Section \ref{sec:fin}, the scheme discussed here exhibits a scaling behaviour for the reevaluation of the underlying Born matrix element proportional to $N^2$, with $N$ being the number of final state particles in the corresponding real emission process. The same scaling behaviour is implicit in the FKS scheme \cite{Frixione:1995ms}, where the mapping of the Born-type matrix element is transferred to the explicit parametrization of phase space for each emitter/ emitted parton pair. In \cite{Frederix:2009yq}, it was shown that within this scheme the scaling behaviour can be reduced to a constant for processes containing symmetric final states. In the following, we want to argue that exactly the same scaling behaviour can be achieved in the scheme discussed here and is indeed implicit in the setup of our scheme, and especially the choice of soft interference terms proposed in Section \ref{sec:softss}. The implementation of this prescription in a numerical code is in the line of future work.\\

The improved scaling behaviour proposed in \cite{Frederix:2009yq} relies on the fact that any $m+1$ phase space can be decomposed into disjoint partitions of phase space that are specified by their behaviour for one of the partons $\ph_i$ becoming soft or collinear to at most one additional parton $\ph_j$: these adjoint pairs are then denoted FKS pairs, where the sum of all subspace partitions reproduces the whole phase space\footnote{Note that the notation between \cite{Frederix:2009yq} and this work differs in the fact that in \cite{Frederix:2009yq}, $\ph_i$ labels the emitted parton that becomes soft or collinear, while in our case this parton is denoted by $\ph_j$. For sake of consistency, we here stick to the notation proposed in \cite{Frederix:2009yq}. }: 
\begin{\eqn*}
\sum_{(i,j)\in \mathcal{P}_\text{FKS}}\,\mathcal{S}_{ij}\,=\,1
\end{\eqn*}
(Eqn (4.16) in \cite{Frederix:2009yq}), where $\mathcal{P}_\text{FKS}$ denotes the set of FKS partitions. Furthermore, the authors observe that for processes displaying symmetries in the final state, which are subsequently reflected in the matrix element and phase space, several partitions $\mathcal{S}_{ij}$ render exactly the same contribution to the final observable, and that therefore the evaluation of at most one of these is sufficient, the others being related by symmetry:
\begin{\eqn}\label{eq:fks_sym}
d\sigma^{(n+1)}(r)\,=\,\sum_{(i,j)\in\overline{\mathcal{P}}_\text{FKS}}\xi_{ij}^{(n+1)}(r)d\sigma_{ij}^{(n+1)}(r),
\end{\eqn}
(eqn (6.7) in \cite{Frederix:2009yq}), where $\xi_{ij}^{(n+1)}(r)$ denotes the process-dependent symmetry factor that relates the total cross section to the one evaluated in the partition denoted by $\mathcal{S}_{ij}$, and $\overline{\mathcal{P}}_\text{FKS}$ now denotes the set of all nonredundant partitions. Note that an important argument here is that all {\sl other} contributions that stem from $\ph_i$ becoming soft, but collinear to a {\sl different} parton $\ph_k$, belong to a {\sl different} partition $\mathcal{S}_{ik}$ and are therefore suppressed via the structure of $\mathcal{S}_{ij}$. Especially for purely gluonic final states, $\overline{\mathcal{P}}_\text{FKS}$ contains only one element.\\

In the scheme discussed here, the subtraction term that reflects the divergences of $\mathcal{S}_{ij}$ is given by Eqn. (\ref{eq:delljtot}):
\begin{\eqn*}
{\mathcal{D}}(\ph_j,\ph_i)\,=\,\mathcal{D}^\text{coll}(\ph_j,\ph_i)\,+\,\delta_{\hat{f}_j,g}\sum_{k\,\neq\,(i,j)}\mathcal{D}^\text{if}(\ph_j,\ph_i,\,\ph_k);
\end{\eqn*}
as discussed in Section \ref{sec:softss}, all contributions from the soft/collinear divergence of $\ph_i,\,\ph_k$ are transferred to the interference term $\mathcal{D}^\text{if}(\ph_k,\ph_i,\,\ph_j)$, corresponding to the singularity structure of a {\sl different} partition, namely $\mathcal{S}_{ik}$. All terms in Eqn. (\ref{eq:delljtot}) come with the same mapping, and, as in the FKS prescription in \cite{Frederix:2009yq}, only the set of nonredundant contributions needs to be evaluated, all others being related by symmetry. Increasing the number of final state gluons then leads to a change in the constant $\xi_{ij}^{(n+1)}(r)$ but does not call for the evaluation of a larger number of nonredundant contributions, as the number of elements in $\overline{\mathcal{P}}_\text{FKS}$ remains constant. Therefore, following this prescription, our scheme equally exhibits a constant scaling behaviour, when the number of gluons in the real emission final state is increased.\\

We finally want to comment that, although the above prescription can naturally lead to a significant improvement in the treatment of real-emission subtractions for multi-parton final states, it is not straightforward to implement in standard Monte Carlo generators that do not internally make use of the symmetries exhibited in Eqn. (\ref{eq:fks_sym}). The implementation of this prescription therefore equally calls for a modification of the NLO tools used for calculating the corresponding process. 
\end{appendix}
\bibliographystyle{utcaps}
\bibliography{../NLO_subtraction}
\end{document}